\documentclass[aps,prl,twocolumn,groupedaddress,notitlepage,%showpacs,
floatfix,superscriptaddress]{revtex4-1}

\pdfoutput=1
\usepackage{graphicx,graphics,epsfig,subfigure,times,bm,bbm,amssymb,amsmath,amsthm,mathrsfs,MnSymbol}
\usepackage{gensymb}
\usepackage{amsfonts}
\usepackage{float}
\usepackage[matrix,frame,arrow]{xypic}
\usepackage[pdftex]{color}
\usepackage{braket}  %Dirac Notation in QM
\usepackage{enumerate}
\usepackage[normalem]{ulem}
\usepackage[usenames,dvipsnames]{xcolor}
\usepackage{multirow}
\usepackage{mathtools}
\usepackage{bbm}
\usepackage{titletoc}

\definecolor{orange}{rgb}{1,0.5,0}
\newcommand{\RNum}[1]{\uppercase\expandafter{\romannumeral #1\relax}}

\newcommand{\ignore}[1]{}
\usepackage{geometry}

\ignore{
	\documentclass[eprintnumbers,amsmath,amssymb,onecolumn,a4paper,caption ]{article}
	\usepackage{amsfonts}
	\usepackage{amssymb}
	\usepackage{mathrsfs}
	\usepackage{mathbbold}
	\usepackage{bbm}
	\usepackage{mathrsfs}
	\usepackage{dcolumn}% Align table columns on decimal point
	\usepackage{bm}% bold math
	\usepackage{times,epsfig,amssymb,amsmath}
	\usepackage{float}
	\usepackage{subfigure}
	\usepackage{geometry}\geometry{left=2.5cm,right=2.5cm,top=3cm,bottom=3cm}

	\usepackage{color}

}

\usepackage{hyperref}
\hypersetup{
	colorlinks=true,       % false: boxed links; true: colored links
	linkcolor=red,          % color of internal links
	citecolor=blue,        % color of links to bibliography
	filecolor=magenta,      % color of file links
	urlcolor=blue,           % color of external links
	runcolor=cyan
}

\begin{document}

	\title{Observation of Bloch Oscillations and Wannier-Stark Localization  on a Superconducting Quantum Processor}
	
	\author{Xue-Yi~Guo}
	\altaffiliation[]{These authors contributed equally to this work.}
	\affiliation{Beijing National laboratory for Condensed Matter Physics, Institute of Physics, Chinese Academy of Sciences, Beijing 100190, China}
	
	\author{Zi-Yong~Ge}
	\altaffiliation[]{These authors contributed equally to this work.}
	\affiliation{Beijing National laboratory for Condensed Matter Physics, Institute of Physics, Chinese Academy of Sciences, Beijing 100190, China}
	\affiliation{School of Physical Sciences, University of Chinese Academy of Sciences, Beijing 100190, China}
			
	\author{Hekang Li}
	\affiliation{Beijing National laboratory for Condensed Matter Physics, Institute of Physics, Chinese Academy of Sciences, Beijing 100190, China}
	\affiliation{School of Physical Sciences, University of Chinese Academy of Sciences, Beijing 100190, China}
	
	\author{Zhan~Wang}
	\affiliation{Beijing National laboratory for Condensed Matter Physics, Institute of Physics, Chinese Academy of Sciences, Beijing 100190, China}
	\affiliation{School of Physical Sciences, University of Chinese Academy of Sciences, Beijing 100190, China}
	
    \author{Yu-Ran~Zhang}
	\affiliation{Theoretical Quantum Physics Laboratory, RIKEN Cluster for Pioneering Research, Wako-shi, Saitama 351-0198, Japan}
	
	\author{Pengtao~Song}
	\affiliation{Beijing National laboratory for Condensed Matter Physics, Institute of Physics, Chinese Academy of Sciences, Beijing 100190, China}
	\affiliation{School of Physical Sciences, University of Chinese Academy of Sciences, Beijing 100190, China}

	\author{Zhongcheng~Xiang}
	\affiliation{Beijing National laboratory for Condensed Matter Physics, Institute of Physics, Chinese Academy of Sciences, Beijing 100190, China}

	\author{Xiaohui~Song}
	\affiliation{Beijing National laboratory for Condensed Matter Physics, Institute of Physics, Chinese Academy of Sciences, Beijing 100190, China}
	
	\author{Yirong~Jin}
	\affiliation{Beijing Academy of Quantum Information Sciences, Beijing 100193, China}

	 \author{Li~Lu}
     \affiliation{Beijing National laboratory for Condensed Matter Physics, Institute of Physics, Chinese Academy of Sciences, Beijing 100190, China}
    \affiliation{School of Physical Sciences, University of Chinese Academy of Sciences, Beijing 100190, China}
    \affiliation{CAS Center for Excellence in Topological Quantum Computation, UCAS, Beijing 100190, China}
    \affiliation{Songshan Lake Materials Laboratory, Dongguan 523808, China}
	
	\author{Kai~Xu}
	\affiliation{Beijing National laboratory for Condensed Matter Physics, Institute of Physics, Chinese Academy of Sciences, Beijing 100190, China}
	\affiliation{CAS Center for Excellence in Topological Quantum Computation, UCAS, Beijing 100190, China}
	
	\author{Dongning~Zheng}
	\email{dzheng@iphy.ac.cn}
	\affiliation{Beijing National laboratory for Condensed Matter Physics, Institute of Physics, Chinese Academy of Sciences, Beijing 100190, China}
	\affiliation{School of Physical Sciences, University of Chinese Academy of Sciences, Beijing 100190, China}
	\affiliation{CAS Center for Excellence in Topological Quantum Computation, UCAS, Beijing 100190, China}
	\affiliation{Songshan Lake Materials Laboratory, Dongguan 523808, China}

	\author{Heng~Fan}
	\email{hfan@iphy.ac.cn}
	\affiliation{Beijing National laboratory for Condensed Matter Physics, Institute of Physics, Chinese Academy of Sciences, Beijing 100190, China}
	\affiliation{School of Physical Sciences, University of Chinese Academy of Sciences, Beijing 100190, China}
	\affiliation{CAS Center for Excellence in Topological Quantum Computation, UCAS, Beijing 100190, China}
	\affiliation{Songshan Lake Materials Laboratory, Dongguan 523808, China}

\begin{abstract}	
The Bloch oscillation (BO) and Wannier-Stark localization (WSL) are fundamental concepts
about metal-insulator transitions in condensed matter physics.
These phenomena have also been observed in semiconductor superlattices and simulated in platforms such as photonic waveguide arrays and cold atoms.
Here, we report experimental investigation of BOs and WSL simulated with a 5-qubit programmable superconducting processor,
of which the effective Hamiltonian is an isotropic $XY$ spin chain.
When applying a linear potential to the system by properly tuning all individual qubits,
we observe that the propagation of a single spin on the chain is suppressed.
It tends to oscillate near the neighborhood of their initial positions, which demonstrates the characteristics of BOs and WSL.
We verify that the WSL length is inversely correlated to the potential gradient.
Benefiting from the precise single-shot simultaneous readout of all qubits in our experiments,
we can also investigate the thermal transport, which requires the joint measurement of more than one qubits.
The experimental results show that, as an essential characteristic for BOs and WSL,
the thermal transport is also blocked under a linear potential.
Our experiment would be scalable to more superconducting qubits for simulating various of out-of-equilibrium problems in quantum many-body systems.
\end{abstract}
	
	\date{\today}
	\maketitle{}
	
\begin{figure*}[t]     	
	\includegraphics[width=0.7\textwidth]{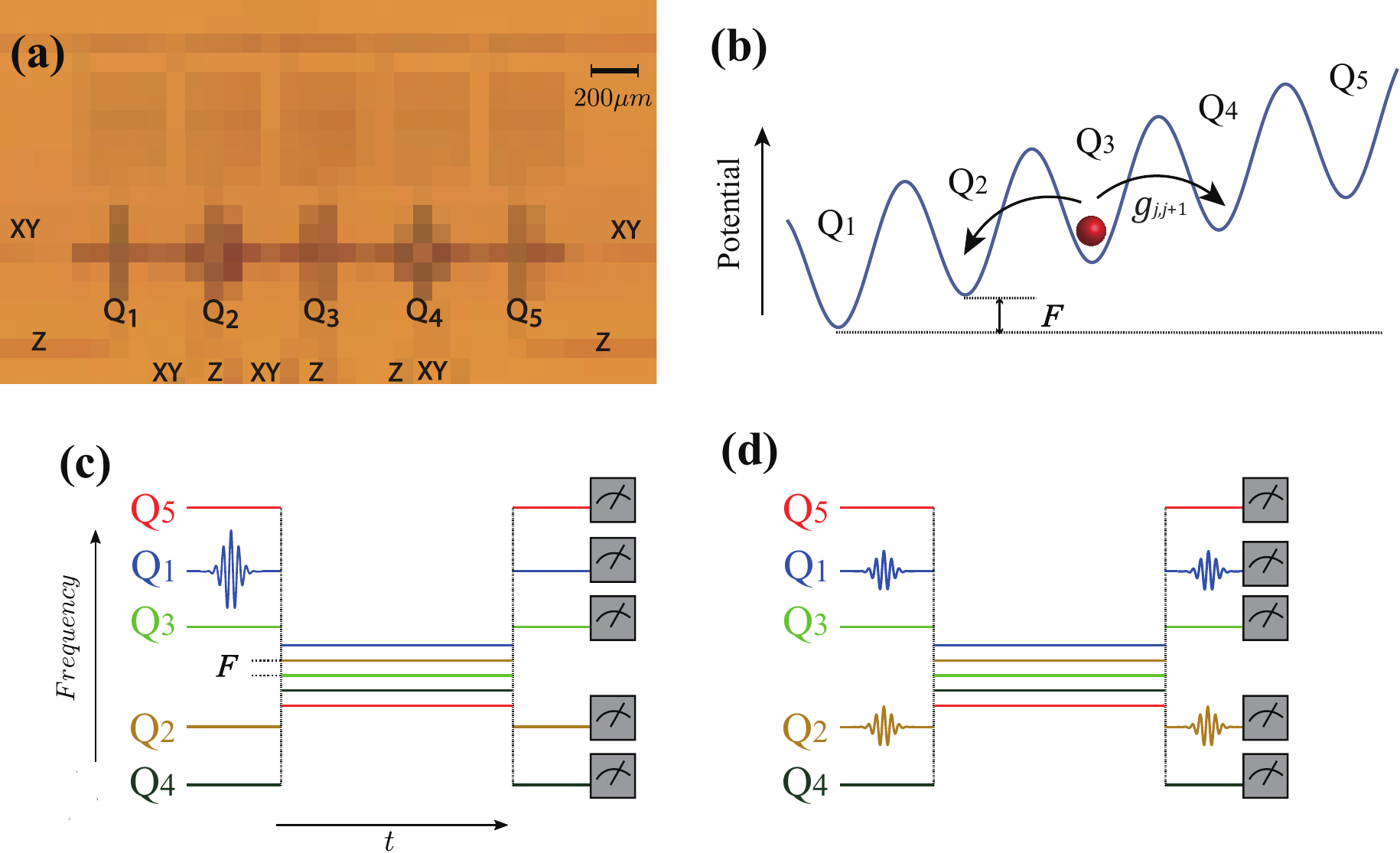}
	\caption{Experimental setup. (a) Circuit diagram of the device. There are 5 superconducting transmon qubits ($Q_1-Q_5$)  arranged into a chain~\cite{You,Koch,Barends3,Barends2013,Mutus,Lucero2}. The nearest-neighbor two qubits are coupled  capacitively, and the negative anharmonicity $U$ of each qubit gives the on-site attractive interaction. The frequency of each qubit is tunable by individual microwave driving through the flux-bias line (Z line). The qubit flipping can be realized by XY lines.
Each qubit is coupled to a resonator for individual and simultaneous readout. More experimental details of the system are presented in the Supplemental Note 1.
(b) The sketch of the corresponding Bose-Hubbard chain with a linear potential. The detuning of the nearest-neighbor two qubits is the potential gradient $F$. The red ball represents the photon (the excitation of the qubit), which can tunnel to the nearest-neighbor sites (black arrows).
(c) Pulse sequences for studying the transport of spin. The qubits are ordered by their frequencies, and initialized at their idle frequencies and the state $\ket{0}$, see Supplemental Note 1. A.  Then, we excite $Q_1$ to the state $\ket{1}$ by a $X$ gate and bias all qubits at the work point with square waves. After the system evolves for a specific time $t$, all qubits are biased back to their idle frequencies, and we finally read out each qubit. (d) Pulse sequences for studying the thermal transport. We use two $X/2$ gates at $Q_1$ and $Q_2$ two prepare the initial state $\ket{X_+X_+000}$, and other sequences is the same as (c).}
\label{fig_1}
\end{figure*}

\noindent
\textbf{\large{Introduction}}\\
The transport phenomena in solids is one of the central topics in condensed matter physics.
About 80 years ago, Bloch and Zener predicted that electrons cannot spread uniformly in a crystal lattice under a constant force, and instead, they would oscillate and localize~\cite{Bloch1929,Zener1934,Wannier1962}.
This oscillation is called Bloch oscillations (BOs), and the corresponding localization is called  Wannier-Stark localization (WSL).
BOs and WSL are typical quantum effects which reveal the wave properties of electrons.
However, they can hardly be observed directly in normal bulk materials due to the requirement of long coherence times.
It is not until the 1990s that these phenomena were observed experimentally in semiconductor superlattices~\cite{Feldmann1992}. Nevertheless, the relaxation time in this type of material is still a bottleneck for studying BOs and WSL.
During the last two decades, the developments in quantum technology have made it possible to simulate these quantum phenomena in artificial quantum systems~\cite{Buluta2009,Georgescu2014}.
Compared with the semiconductor superlattice systems, these artificial quantum systems have much longer decoherence times making  them  suitable for the experimental study of BOs.  BOs in  bosonic systems have been observed in the cold atoms~\cite{Dahan1996,Anderson1998, Morsch2001, Fattori2008, Gustavsson2008, Alberti2009, Haller2010, Meinert2014, Preiss2015, Geiger2018} and photonic waveguide arrays~\cite{Morandotti1999}, etc.

Due to the scalability, long decoherence time and high-precision control,
the superconducting circuit~\cite{Makhlin2001,Gu2017} has become a competitive candidate  for achieving universal quantum computation
and have demonstrated quantum supremacy~\cite{Arute2019}.
Superconducting circuits can be fabricated into different lattice structures, such as 1D chain, ladder, fully connected graphs, and 2D square lattice.
It is  a versatile platform for performing various kinds of quantum-simulation experiments, e.g., quantum many-body dynamics~\cite{Eisert2015,Braumuller2017,Xu2018,Roushan,Salathe,Barends,Zhong,Song1,Flurin,Ma2019,Yan2019,Ye2019,Guo2019,Xu2019}, quantum chemistry~\cite{OMalley,Kandala},
and implementing quantum algorithms~\cite{Lucero,Gong,Barends2,Zheng,Song2,Song3}.
Our quantum processor with 1D array of superconducting qubits is well suited for studying essential transport properties of spin and energy in BOs and WSL.
Remarkably, measurements of energy transport are absent in previous simulations,
which needs capability of multiqubits single-shot simultaneous readout in obtaining nearest-neighbor two-site correlations.

In this work, we experimentally investigate BOs and WSL of spin system  on a 5-qubit superconducting processor.
The effective Hamiltonian can be described by an  isotropic $XY$ chain.
By manipulating the frequencies of superconducting qubits precisely, we can construct a linear  potential. Under this type of potential,  we observe that the spin can hardly propagate through the lattice during the quench dynamics. It tends to oscillate at the vicinity of initial positions, which is a typical phenomenon of  BOs and WSL.
In addition, using the maximum probability of a photon propagating from one boundary to another to represent the  WSL length, we can demonstrate that the localization length is inversely correlated to the potential gradient.
By performing precise simultaneous readout of two superconducting qubits, we can also study the thermal transport of the system. It is shown that the energy transport is suppressed as well by the linear potential.
\\

\noindent
\textbf{\large{Results}}\\
\textbf{Experimental setup and model.}
In this experiment, our superconducting processor contains 5 qubits arranged into a 1D chain, with the capability of high-precision simultaneous readouts and full controls, see Fig.~\ref{fig_1}(a).
%The decoherence time of the superconducting qubits is more than $17$~$\mu$s (see Supplemental Note 1.~A).
The Hamiltonian of the system can be described by the 1D Bose-Hubbard model, which reads($\hbar=1$)~\cite{Roushan,Yan2019,Ye2019}
\begin{equation} \label{Hbh}
\hat H =\sum_{j=1}^4g_{j,j+1}(\hat{a}^\dag_{j} \hat{a}_{j+1}+\hat{a}^\dag_{j+1} \hat{a}_{j})+\sum_{j=1}^{5}(\frac{U_j}{2}\hat{n}_j(\hat{n}_{j}-1)+h_{j}\hat{n}_{j}),\\
\end{equation}
where $\hat{a}_{j}^\dagger$ ($\hat{a}_{j}$) is the photon creation (annihilation) operator,  $\hat{n}_{j} \equiv \hat{a}^\dagger_{j} \hat{a}_ {j}$ is the number operator,
$g_{j,j+1}$ is the nearest-neighbor coupling strength,
$U_j<0$ is the on-site attractive interaction resulted from the anharmonicity,
and $h_j$ is the local  potential which is tunable by DC biases through $Z$ lines.
To realize BOs, we let $h_j$ vary linearly along the lattice sites, i.e, $h_j = F j$,
where $F$ is the potential gradient or the detuning of  nearest-neighbor two qubits, see Fig.~\ref{fig_1}(b).

\begin{figure*}[t]     	
	\includegraphics[width=0.95\textwidth]{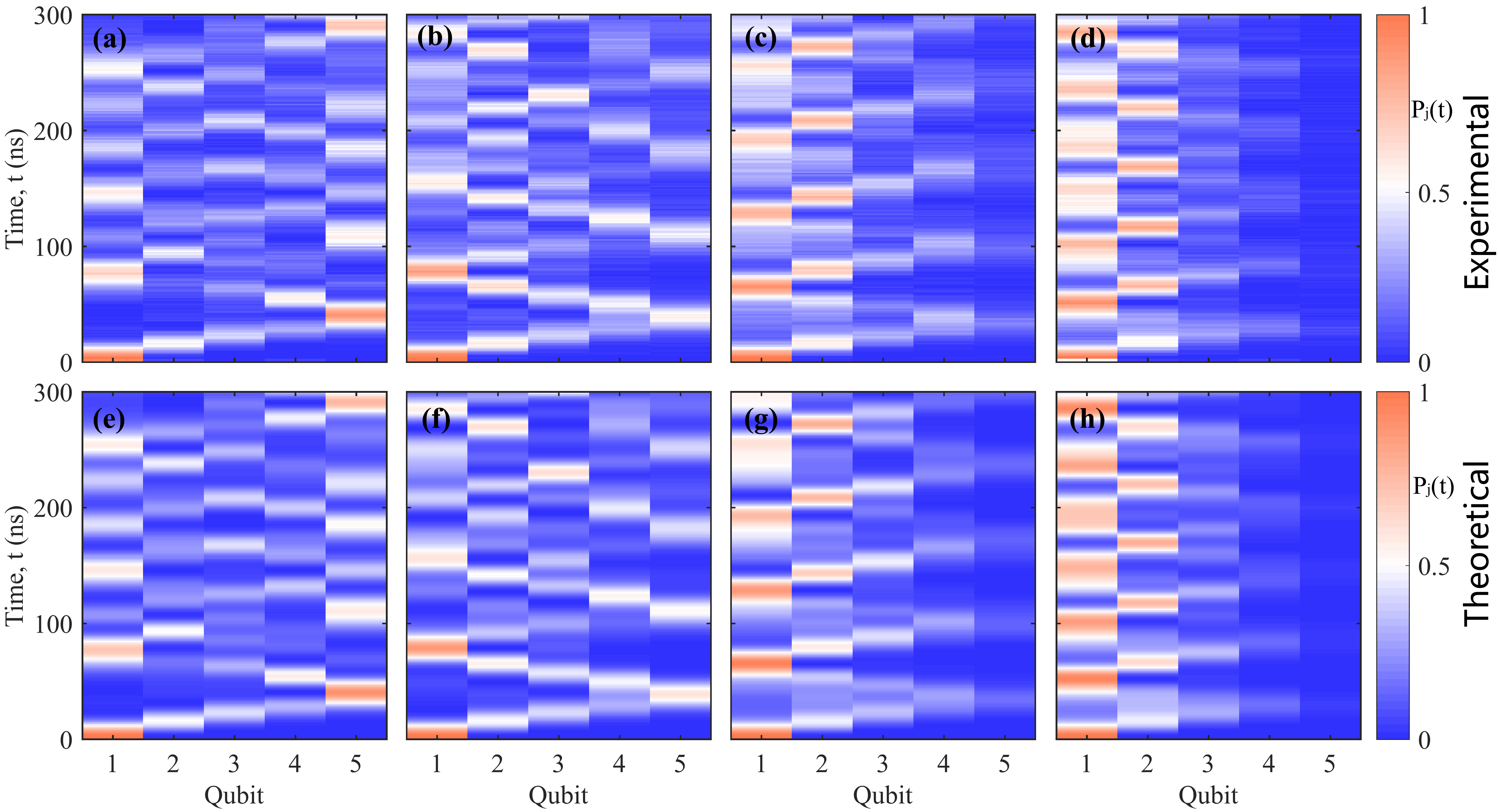}
	\caption{Spin transport. (a--d) The experimental results for the time evolution of density distribution $P_j(t)$ up to $300$~ns. The initial state is $\ket{10000}$, and the potential gradient (a) $F/2\pi=0$, (b) $F/2\pi=5$, (c) $F/2\pi=10$,(d) $F/2\pi=15$~MHz.
		The photon or the spin can exhibit a light-cone-like spreading in the system.
		As the increase of $F$, the  photon becomes more difficult to propagate to the right boundary, instead, it tends to oscillate in the vicinity of the initial position.
Each point shows the average of $6\times 100$ single-shot measurements.
		(e--h) The corresponding theoretical results of (a--d). The numerical and experimental results are consistent with each other. The details of numerical method are presented in \textbf{Methods}.}
	\label{fig_2}
\end{figure*}

 In this superconducting circuits, since  $|U_j|/g_{i,j} \gg 1$ and $U_j$ is staggered to suppress higher order tunneling (see Supplemental Note 1.~A), the Fock space of
 the photons at each qubit can be truncated to two dimensions. Thus, the model is equivalent to a spin-$\frac{1}{2}$ system, and the nonlinear term can be neglected.
 Therefore, the effective Hamiltonian of Eq.~(\ref{Hbh}) can be reduced to an isotropic $XY$ model~\cite{Yan2019,Ye2019}
 \begin{align} \label{Hxy}
  \hat H_{\textrm{eff}}=\sum_{j=1}^4g_{j,j+1}(\hat{\sigma}^+_{j} \hat{\sigma}^-_{j+1}+\hat{\sigma}^+_{j+1} \hat{\sigma}^-_{j})+\sum_{j=1}^5h_{j}\hat{\sigma}^+_{j}\hat{\sigma}^-_{j} ,
 \end{align}
 where $\hat{\sigma}^{\pm} = (\hat{\sigma}^x\pm i\hat{\sigma}^y)/2$, and $\hat{\sigma}^{x,y,z}$ are Pauli matrices.
 According to Eqs.~(\ref{Hbh}--\ref{Hxy}), we know that the system has an $U(1)$ symmetry, so that the total spins $\sum_{j=1}^5\hat{\sigma}_j^+\hat{\sigma}^-_{j}$ for $\hat H_{\textrm{eff}}$ (or the total photon number $\sum_{j=1}^5\hat{n}_{j}$ for $\hat H$)  are conserved.
 In the following discussion, we do not distinguish the photons and spins.
 In addition, this system is time-independent, thus the energy is also conserved, where we can explore both spin and thermal transport in this system.
 Note that we only consider the evolution time $t\leq300$~ns in the experiment, which  is much smaller than decoherence time  (more than $17$~$\mu$s, see Supplemental Note 1.~A). Therefore, the above conservation laws are nearly unbroken under the impact of decoherence.
\\

\noindent
\textbf{Spin transport.}
Firstly, we study the spin transport after a quantum quench.
The explicit experiment sequences are shown in Fig.~\ref{fig_1}(c).
We initially excite the leftmost qubit $Q_1$ from the state $|0\rangle$ to $|1\rangle$ by a $X$ gate, i.e., the initial state is $|\psi(0)\rangle=|10000\rangle$.
Then, each qubit is biased to the working frequency with the fast $Z$ pulse, and the system will evolve under the Hamiltonian (\ref{Hxy}).
Finally, we measure, for each qubit, the probability distribution of state $|1\rangle$, i.e., the density distribution of the photon or spin, defined as
 \begin{align}
 P_j(t):= \langle\psi(t)|\hat{\sigma}^+_{j}\hat{\sigma}^-_{j} |\psi(t)\rangle,
\end{align}
where $|\psi(t)\rangle=e^{-i\hat Ht}|\psi(0)\rangle$ is the wave function of the system at time $t$.
As shown in Fig.~\ref{fig_2}(a), when $F=0$, the spin displays a light-cone-like propagation without any restrictions
and can exhibit a reflection when approaching the boundaries~\cite{Yan2019,Ye2019}.
Nevertheless, according to Figs.~\ref{fig_2}(b--d), when $F\neq 0$, the spin transport is blocked.
With an increase of $|F|$, the spin can hardly propagate from the leftmost to the rightmost.
Instead, it tends to oscillate around the neighbor of the initial position,
and this is a typical signature of BOs and WSL.
In Figs.~\ref{fig_2}(e--h), we present the corresponding numerical results, which are consistent with the experimental results.
From Fig.~\ref{fig_2}(d), we can know that the BO frequency is about $50$~ns when $F/2\pi=15$~MHz, which is much smaller than the decoherence time of the superconducting qubits.

\begin{figure}[t]     	
	\includegraphics[width=0.48\textwidth]{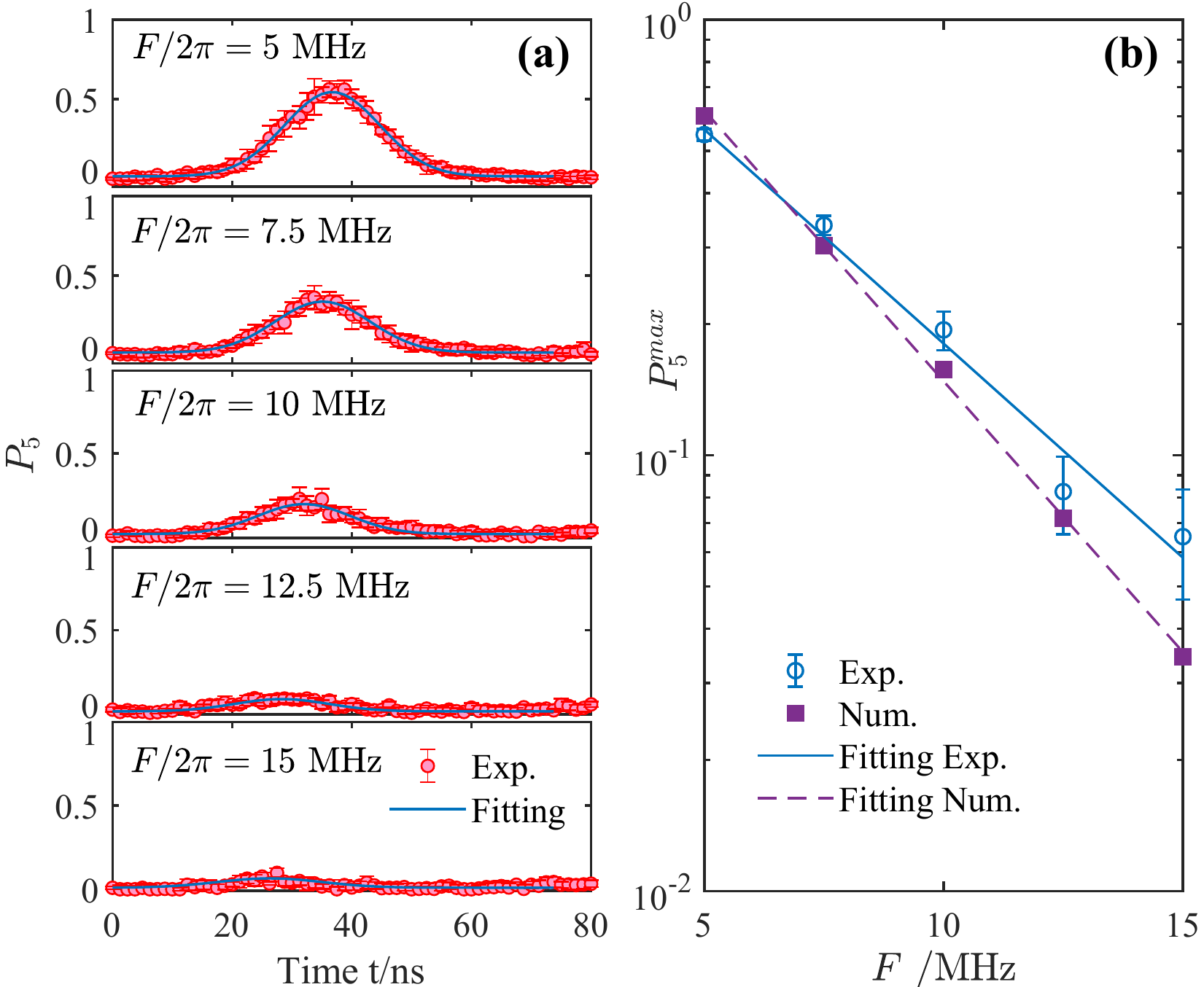}
	\caption{(a) The time evolution of photon occupancy probabilities at $Q_5$ for different potential gradients $F$. The initial state is $\ket{10000}$.
	The red solid circles are experimental data points, and the blue lines are Gaussian fittings. The estimate method of errors is presented in \textbf{Methods}.
 (b) The  relation between $P_5^{\text{max}}$ and $F$, where the experimental $P_5^{\text{max}}$ is the peak value of Gaussian curve shown in (a). The corresponding error bars are the fitting errors. The numerical data points are peak values of the first wavefronts of $P_5(t)$ without Gaussian fitting~\cite{Yan2019}. Here, $\ln P_5^{\text{max}}$ is nearly linearly related to the potential gradients $F$. The solid and dashed lines are the corresponding linear fittings of the experimental and theoretic results, respectively.}
	\label{fig_3}
\end{figure}

\begin{figure}[t]     	
	\includegraphics[width=0.45\textwidth]{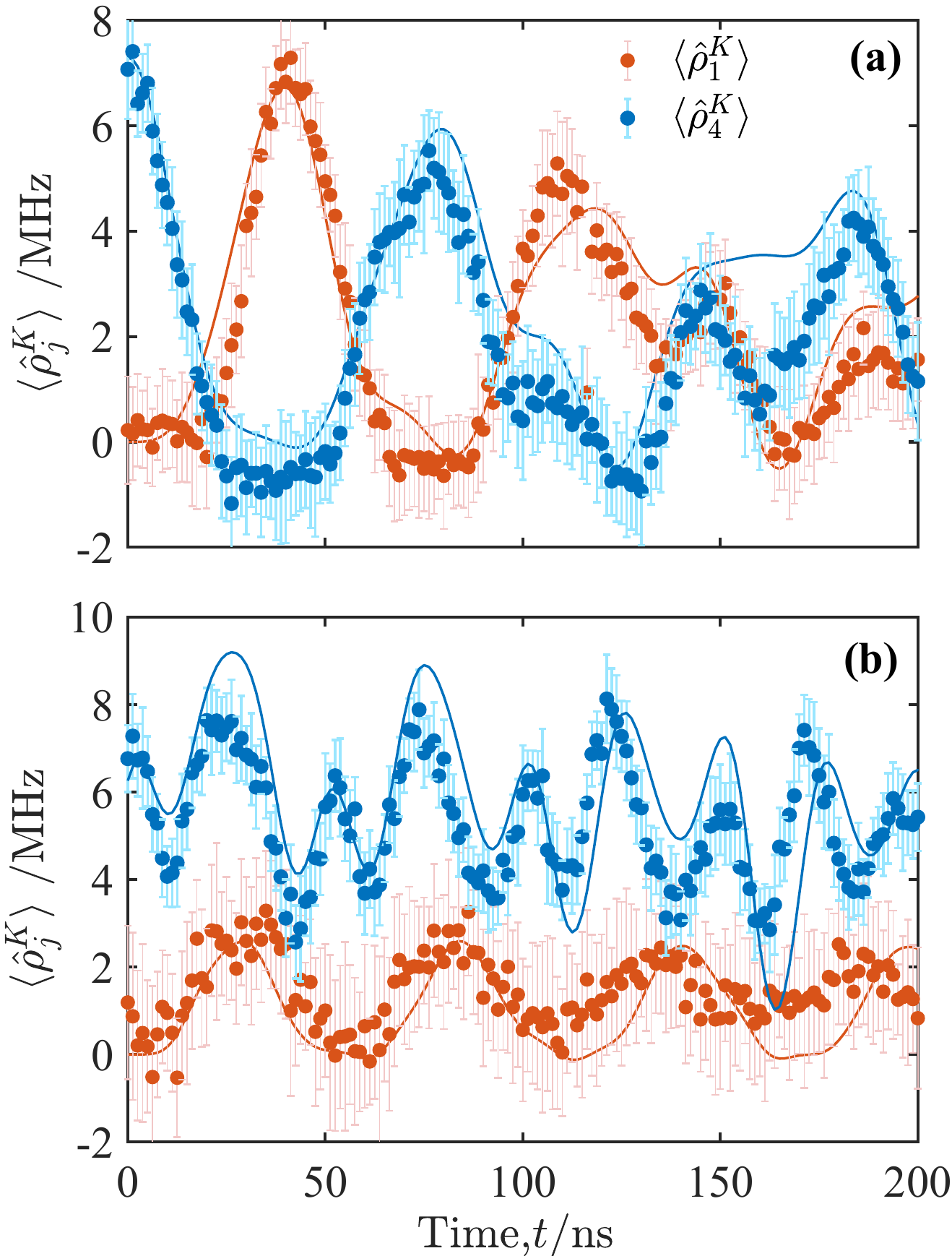}
	\caption{Time evolution of kinetic energy densities at two edges, i.e., $\langle \hat\rho^{K}_1(t)\rangle$ and $ \langle \hat \rho^{K}_4(t)\rangle$. Symbols are experimental data points, and the solid lines represent numerical results, where the decoherence and dephase are considered.  The initial state is $|X_+X_+000\rangle$.  Each point shows the average of $10\times 200$ single-shot measurements, and the estimate method of errors is presented in \textbf{Methods}. (a) $F/2\pi=0$~MHz. The energy densities $\langle \hat\rho^{K}_1\rangle$ and $ \langle \hat \rho^{\text{E}}_4\rangle$ can almost exchange into each other indicating that the energy transport is free in this case.
(b) $F/2\pi=15$~MHz. There is no crossing between the curves $\langle \hat\rho^{K}_1(t)\rangle$ and $ \langle \hat \rho^{K}_4(t)\rangle$. The existence of this energy gradient between two edges shows that the energy transport is compressed. That is, the energies at one side can hardly spread to the other side.}
	\label{fig_4}
\end{figure}

Now we extract the oscillation amplitude or WSL length $\xi_{WS}$.
Generally,  in the presence of linear potential, the single-particle wave function is localized and has the form $\psi(x)\sim Ae^{-x/\xi_{WS}}$, where $A$ is a normalized factor.
Hence, the probability that a particle can propagate the distance $r$, i.e., $P(r)$, should satisfy $P(r)\propto e^{-r/\xi_{WS}}$.
In this experiment, the existence of boundaries makes it challenging to obtain the localization length.
To overcome this difficulty, we propose another method to extract the WSL length.
With the maximum photon occupancy probabilities at $Q_5$, defined as $P_5^{\text{max}} := \max_{t>0} P_5(t)$,
we can obtain the WSL length using $\xi_{WS}\propto 1/\ln P_5^{\text{max}} $.
In Supplemental Note 2, we present a phenomenological proving of this relation.
To extract more reliable $P_5^{\text{max}}$, we use Gaussian function to fit $P_5(t)$ and take the corresponding peak value as  $P_5^{\text{max}}$, see Fig.~\ref{fig_3}(a).
Now we study the relation between the potential gradient $F$ and $P_5^{\text{max}}$.
For a WSL system, the localization length $\xi_{WS}$
is inversely proportional to $F$, i.e., $\xi_{WS} \propto 1/F$.
Hence, we expect that $\ln P_5^{\text{max}} \propto F$.
According  to Fig.~\ref{fig_3}(b), we can find that both the numerical simulation  and experimental results are consistence with this relation.
\\

\noindent
\textbf{Thermal transport.}
Now we focus on the thermal transport in this system.
For a 1D chain, the energy density at the $j$-th bond is defined as $\hat \rho^{E}_j= \hat H_{j,j+1}\equiv\hat \rho^{K}_j + \hat \rho^{P}_j$, where $\hat \rho^{K}_j$ and $\hat \rho^{P}_j$ denote
\textit{kinetic energy} and \textit{potential energy} densities, respectively. From Eq.~(\ref{Hxy}), these two quantities can be expressed as
\begin{eqnarray} \label{ed} \nonumber
&&\hat \rho^{K}_j=\frac{1}{2}g_{j,j+1}(\hat{\sigma}^x_{j} \hat{\sigma}^x_{j+1}+\hat{\sigma}^y_{j} \hat{\sigma}^y_{j+1}),\\
&&\hat \rho^{P}_j = h_{n}\hat{\sigma}^+_{j}\hat{\sigma}^-_{j}
 +h_{j+1}\hat{\sigma}^+_{j+1}\hat{\sigma}^-_{j+1}.
\end{eqnarray}
In general, the thermal transport is closely related to the electronic charge transport in a classical metal system,
which is known as the Wiedemann-Franz~\cite{Franz,Chester} law,
 i.e., $\lambda/\sigma = L T$, where  $\lambda $ is thermal conductance, $\sigma $ is electronic conductance, $T$ is temperature, and $L$ is Lorenz number.
Eq.~(\ref{ed}) shows that the potential energy only depends on the spin distribution, which displays BOs and WSL as discussed in the previous section.
Here, we consider the time evolution of the kinetic energy density.

In Fig.~\ref{fig_1}(d), the pulse sequences of this experiment are presented .
To study the transport of $\hat \rho^{K}_j$,  the kinetic energy densities should exist a gradient between two edges at the  initial state.
Here, we choose the initial state as $\ket{X_+X_+000}$, where $\ket{X_+}=\frac{1}{\sqrt{2}}(\ket{0}+\ket{1})$ is the eigenstate of $\hat \sigma^x$ with eigenvalue $1$ and can be prepared by $X/2$ gate.
We can verify that, with this initial state, the kinetic energy at left edge is larger than one at right edge, so this initial state can be used to study the thermal transport.
Then,  $\langle \hat \rho^{K}_1(t)\rangle$ and $\langle\hat\rho^{K}_4(t)\rangle$,  i.e., the kinetic energy densities of two edges, are measured, where the simultaneous readout of the nearest-neighbor two qubits is necessary.
As shown in Fig.~\ref{fig_4}(a), when $F = 0$, the kinetic energies of two edges can exchange almost freely.
Nevertheless, from Fig.~\ref{fig_4}(b), we can find that the difference between $\langle\hat\rho^{K}_1(t)\rangle$ and $\langle\hat\rho^{K}_4(t)\rangle$ always exist, when $F/2\pi=15$~MHz.
Therefore, similar to the spins, the thermal transport is also suppressed under the linear potential.

Due to the $U(1)$ symmetry, the quench dynamics can be decomposed into different particle-number subspace, and different subspaces are decoupled with each other. For the initial state $\ket{X_+X_+000}$, the photons only bunch at $Q_1$ or $Q_2$.
Despite existence of two-excitation populating for this initial state, Hamiltonian (\ref{Hxy}) can still effectively describe the dynamics of this system, since two excitations can hardly bunch at a same site due to large and staggered $U_j$.
Thus, we can use Slater determinant to calculate the dynamics of two-excitation sector.
We can verify that  the spins are localized among all of these subspaces with $F\neq0$.
The spins can hardly propagate to the other side, so  $Q_4$ and $Q_5$ almost remain at the initial state $|00\rangle$.
Therefore, the change of $\langle\hat\rho^{K}_4\rangle$ is small in this case, i.e., the kinetic energy can hardly transport from the left edge to the right edge.
In this picture, we can know that the restriction of energy transport origins from the localization of the spins, which is identified with the classical Wiedemann-Franz  law.
\\

\noindent
\textbf{\large{Discussion}}\\
In summary, we have reported the experimental observation of BOs and WSL on a 5-qubit superconducting processor.
We provide another representation of the WSL length for a finite size system, i.e., the probability that a photon can propagate from one edge to anther edge. Using this representation, we verify that the WSL length is inversely proportional to the potential gradients.
Furthermore, benefiting from the precise simultaneous readout of two qubits, the thermal transport in this system is also studied.
The evolution of the energy densities shows that the thermal transport, akin to the spins, is not free under the linear potential, neither.

Comparing to the other artificial quantum many-body systems, one of the most significant advantages of the superconducting quantum circuits is that the states of superconducting qubits can be measured in an arbitrary basis. Thus, it enables us to study the thermal transport associated with BOs, which is generally a challenge for other platforms.
Our results reveal that the superconducting quantum circuits can be considered as alternative synthetic quantum systems for experimentally exploring  BOs and other quantum physics.
Our platform may be useful for the further study of BOs, such as studying the BO frequency and spin current (see Supplemental Note 3), and imaging the Bloch band through BOs~\cite{Geiger2018}.
Our platform can also be extended to studying the transport phenomenon in other specific systems, for instance, in the presence of disorder potentials or engineered noises. In addition, it is meaningful to extend this system to the interacting case, and the Stark many-body localization may be realized in this system~\cite{Schulz2019,Nieuwenburga2019}.
To explore these problems, our system could be scaled to include more qubits with longer decoherence time.
\\

\noindent
\textbf{\large{Methods}} \\
\textbf{Setup.}
This 5-qubit device is made in the following  processes: $i$) Depositing Aluminum. A 100-nm-thick Al layer is deposited on a $10\times10$~mm c-plane sapphire substrate by means of electron-beam evaporation with a base pressure lower than $10^{-9}$ Torr. $ii$) Etching the wires, resonators, and capacitor.  We use a direct laser writer (DWL66+) and wet etching to produce microwave coplanar waveguide resonators, transmission lines, control lines, and capacitors of the Xmon qubit. The resist used here is S1813, and wet-etching process is carried out with Aluminum Etchant Type A. $iii$) Fabricating Josephson junctions. The Josephson junctions of qubits are fabricated by the double-angle evaporation process. In this step, the undercut structure is made by a PMMA-MMA double layer EBL resist following the process similar to one reported in Ref.~\cite{Barends2013}. During the evaporation, the bottom electrode is about 30~nm thick, while the top electrode is about 100 nm thick with intermediate oxidation.

We package the device in an aluminum alloy sample box and fix the box on the mixing chamber stage of a dilution refrigerator. The temperature of the mixing chamber is below 15 mK during measurements. In order to reduce the external electromagnetic interference, an aluminum can and a $\mu$-metal can are placed outside the sample box.
		
For each qubit, microwave pulses are applied through XY lines to rotate the qubit state between $|0\rangle$ and $|1\rangle$. Such XY pulses are formed by modulating continuous microwave signals sent from arbitrary waveform generators (AWGs: Zurich instruments HDAWG) via IQ mixers. To control all 5 qubits, the signal from a microwave source is divided into 5 channels through a power splitter, and each channel is amplified by a 11 dBm level. Current pulses are applied through Z control lines to tune the qubit frequencies.  We use a DC current source (Yokogawa GS220) to apply static direct current to bias a qubit to its idle frequency and use an AWG to apply a fast current pulse to tune the qubit frequencies dynamically. Such static direct current and fast current pulses are combined by a bias-Tee, of which the capacitor is removed.

Readout pulses are composed of five tones at 40-MHz intervals. Each pulse corresponds to one qubit and is applied through the readout line. The output signals are amplified by a broad band Josephson parametric amplifier (JPA)~\cite{Mutus}  and a low temperature HEMT amplifier before further enhancement by a room temperature amplifier. The amplified signal is demodulated by a IQ Mixer and acquired by an analog-digital converter (ADC: Alazar ATS9360).

Attenuators, filters and isolators are used to reduce and  isolate the noise from the electronic instruments, active electronic components (such as JPA and HEMT) and passive components outside the mixing chamber.
\\

\noindent
\textbf{Error estimation.}
In our experiments, for the single-qubit readout, e.g., Fig.~\ref{fig_3}(a), each point shows the average of $6\times100$ single-shot measurements. To estimate the errors, we equally divide these single-shot readout data into 6 groups (each group contains 100
readouts). Thus, we can obtain 6 expectation values for each point, and the
error bar is the standard deviation of these 6 expectation values.
For the two-qubit readout, e.g., Fig.~\ref{fig_4}, each point shows the average of $10\times200$  single-shot measurements.
We use the same method to estimate the errors, where the readout data are  equally divided into 10 groups.
\\

\noindent
\textbf{Numerical methods.}
The numerical results are obtained by numerically solving the Lindblad master equation, which reads
\begin{eqnarray} \label{ed} \nonumber
\frac{d}{dt}\hat \rho(t) = &&-i[\hat H, \hat\rho(t) ] \\ \nonumber
+ &&\frac{1}{2}\sum_{n=1}^5\big[2\hat\Gamma_n\hat\rho(t) \hat\Gamma_n^\dagger
 - \hat\rho(t) \hat\Gamma_n \hat\Gamma_n^\dagger - \hat\Gamma_n \hat\Gamma_n^\dagger\hat\rho(t)\big]\\ \nonumber
+&& \frac{1}{2}\sum_{n=1}^5\big[2\hat A_n\hat\rho(t) \hat A_n^\dagger
 - \hat\rho(t) \hat A_n \hat A_n^\dagger - \hat A_n \hat A_n^\dagger\hat\rho(t)\big] ,\\
\end{eqnarray}
where $\hat\rho(t)$ is the density matrix at time $t$, and Lindblad operators $\hat\Gamma_n=\sqrt{1/T_1}\hat a_n$ and $\hat A_n = \sqrt{1/T_2^*}\hat a_n^\dagger\hat a_n$ represent the excitation leakage and dephasing, respectively.
The corresponding parameters applied here have been calibrated experimentally, and the details are shown in Supplemental Note 1. A.
\\

\noindent
\textbf{\large{Data availability}} \\	
All data not included in the paper are available upon reasonable request from the
corresponding authors.\\

\noindent
\textbf{\large{Acknowledgements}} \\
This work was supported by NSFC (Grant
Nos. 11774406, 11934018, 11904393), National Key R \& D Program of China
(Grant Nos. 2016YFA0302104, 2016YFA0300600, and 2017YFA0304300),
Strategic Priority Research Program of Chinese
Academy of Sciences (Grant No. XDB28000000),
Japan Society for the Promotion of Science (JSPS) Postdoctoral Fellowship (Grant No. P19326), and the JSPS KAKENHI (Grant No. JP19F19326).
\\

\noindent
\textbf{\large{Author contributions}} \\
Z. Y. G., H. F. and D. Z. conceived the idea, X. Y. G. performed the
experiments with assistances from Z. W., P. T. S. and K. X.,
H. K. L. fabricated the device with assistances from Z. C. X., X. H. S.,
L. L. and Y. R. J.,
Z. Y. G. performed the calculations with the help of Y. R. Z.,
X. Y. G., Z. Y. G., H. F. and D. Z. co-wrote the paper with comments from
all co-authors.
\\
	
\noindent
\textbf{\large{Competing interests}} \\	
The authors declare no competing  interests.

	%%%%%%%%%%%%%%%%%%%%%%%%%%%%

	%%%%%%%%%%%%%%%%%%%%%%%%%%%%%%%%%
	%%%%%%%%%%%%%%%%%%%%%%%%%%%%%%%%%
	%

%%%%%%%%%% Merge with supplemental materials %%%%%%%%%%
\clearpage \widetext
\begin{center}
	\section{Supplemental Material\\ \textit{Observation of Bloch Oscillations and  Wannier-Stark Localization on a Superconducting Processor}}
\end{center}
%%%%%%%%%% Prefix a "S" to all equations, figures, tables and reset the counter %%%%%%%%%%
\setcounter{equation}{0} \setcounter{figure}{0}
\setcounter{table}{0} \setcounter{page}{1} \setcounter{secnumdepth}{3} \makeatletter
\renewcommand{\theequation}{S\arabic{equation}}
\renewcommand{\thefigure}{S\arabic{figure}}
\renewcommand{\bibnumfmt}[1]{[S#1]}
\renewcommand{\citenumfont}[1]{S#1}
%\renewcommand\thesection{S\arabic{section}}
%%%%%%%%%% Prefix a "S" to all equations, figures, tables and reset the counter %%%%%%%%%%

\makeatletter
\def\@hangfrom@section#1#2#3{\@hangfrom{#1#2#3}}
\makeatother

%---------------------------------------------------------------------------

	\maketitle

		This Supplementary Information contains  details of the experiment  including: the experimental setup,  amplification performance of the Josephson parametric amplifier (JPA), characterization of frequency multiplexed readout, decoherence time of each qubit, coupling strength between nearest neighbor qubits,  Z control line crosstalk calibrations, delay time calibration for all control channels, square pulse distortion corrections, preparation of initial states like $|X_+X_+000\rangle$, and calibration of dynamical phase induced dune to frequency tuning. Finally, we present a phenomenological analysis about the Wannier-Stark localization length in a finite-size system, and give a numerical result of the dynamics of spin currents.
		
\clearpage

		\begin{table}
			\centering
			\scriptsize
			\renewcommand\arraystretch{1.6}
			\resizebox{0.6\textwidth}{!}{
				\begin{tabular}{c |c c c c c }
					\hline
					\hline
					
					~ &$Q_1$  &$Q_2$  &$Q_3$  &$Q_4$  &$Q_5$\\
					\hline
					$\omega_i^0/2\pi$~(GHz)
					&$5.502$  &$4.999$  &$5.433$  &$4.968$  &$5.531$\\			
					$\omega_i/2\pi$~(GHz)  &$5.361$  &$4.983$  &$5.327$  &$4.868$  &$5.457$\\			
					$U/2\pi$~(MHz)  &$-242$  &$-196$  &$-239$  &$-196$  &$-242$\\
					$T_{1,i}$~($\mu$s)  &$17$  &$30$  &$42$  &$17$  &$36$\\
					$T_{2,i}^*$~($\mu$s)  &$1.53$  &$4.39$  &$2.20$  &$2.19$  &$2.25$\\
					\hline
					%					\hline
					%					&$Q_1Q_2$  &$Q_2Q_3$  &$Q_3Q_4$  &$Q_4Q_5$\\
					$g_{12}/2\pi$~(MHz) &\multicolumn{2}{c}{$14.60$} &~ &~  &~    \\
					$g_{23}/2\pi$~(MHz) &~& \multicolumn{2}{c}{$14.65$}  &~  &  ~    \\
					$g_{34}/2\pi$~(MHz) &~&~& \multicolumn{2}{c}{$14.17$}  &~     \\
					$g_{45}/2\pi$~(MHz) &~&~&~& \multicolumn{2}{c}{$14.26$}    \\
					\hline
					$\omega^r_i/2\pi$~(GHz)  &$6.766$  &$6.7266$  &$6.687$  &$6.654$  &$6.612$\\
					$F_{0,i}$  &$0.981$  &$0.957$  &$0.957$  &$0.923$  &$0.971$\\
					$F_{1,i}$  &$0.853$  &$0.897$  &$0.891$  &$0.859$  &$0.917$\\

				\end{tabular}
				
			}
			\caption{Qubit characteristics.
				$\omega_{\textrm{i}}^{\textrm{0}}$ is the zero flux biased frequency of $Q_i$. $\omega_{\textrm{i}}$ is the idle
				frequency of $Q_i$. $\omega^{\textrm{r}}_i$ is the readout resonator frequency of $Q_i$. $T_{1,i}$ is the energy relaxation time of $Q_i$ at the idle frequency. $T_{2,i}^*$ is the dephasing time of $Q_i$ at the idle frequency. $U/2\pi$ is the non linearity ($f_{21}-f_{10}$) of $Q_i$ measured at the zero flux bias. $F_{1,i}$ ($F_{0,i}$) is the measured probability of $|1\rangle$ ($|0\rangle$) when $Q_i$ is prepared in $|1\rangle$ ($|0\rangle$). $g_{i,i+1}$ is the coupling strength between $Q_i$ and $Q_{i+1}$.}
			\label{tab1}	
		\end{table}

\section{Supplementary Note 1. Details of the experiment}
		
		\subsection{Experimental setup}
		\begin{figure}[!htp]
			\centering
			\includegraphics[width = 0.9\linewidth]{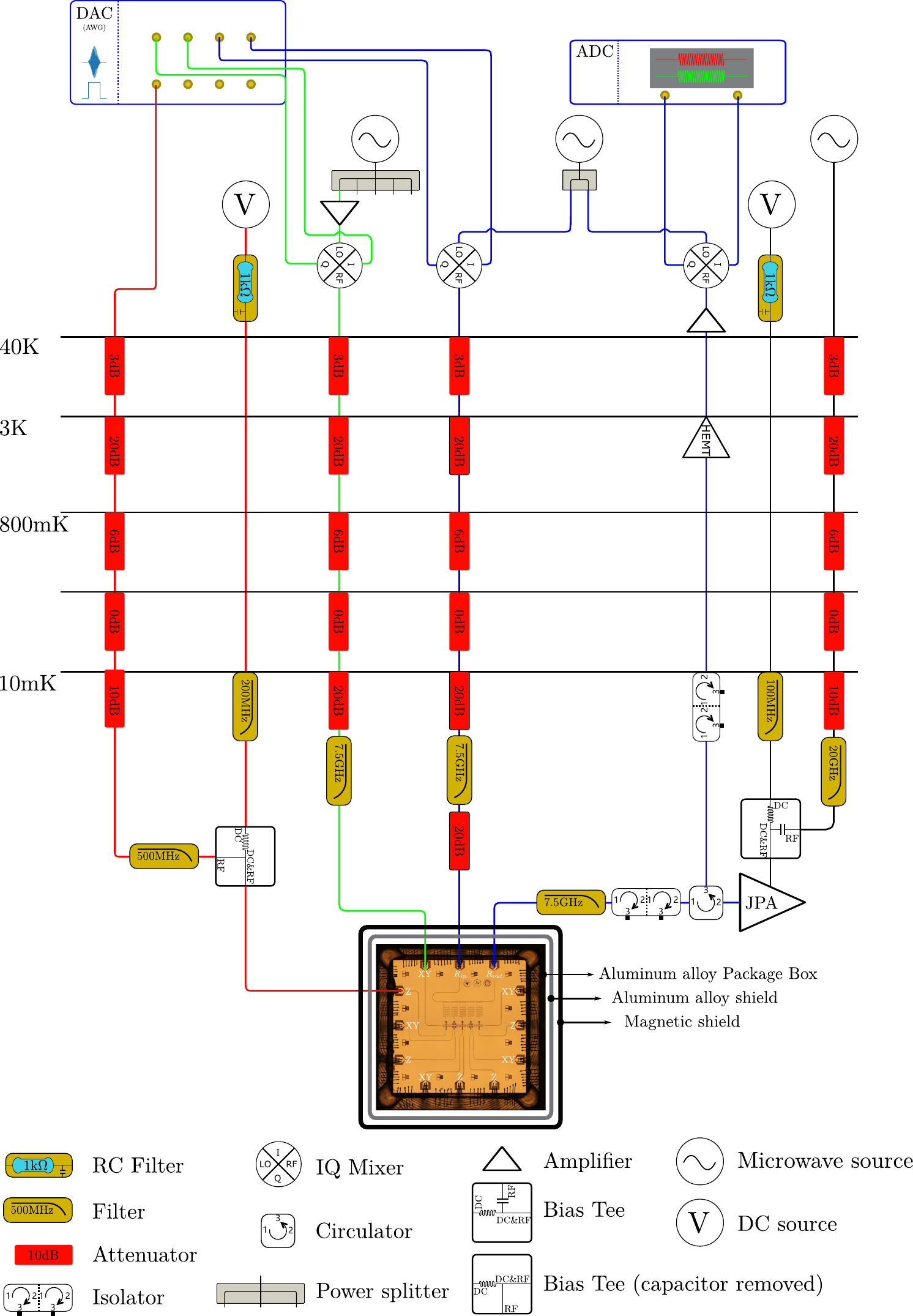}
			\caption{Schematic diagram of the measurement system for superconducting quantum chip.}\label{expsetup}
		\end{figure}

		The schematic diagram of our experimental setup is shown in Supplementary Figure~\ref{expsetup}.		
        The measurement platform contains XY lines ( green line),  Z control lines (red line), and readout line (blue line).
        The more details of measurement platform is presented in the \textbf{Methods}.
		The device photo is shown on the bottom of Supplementary Figure~\ref{expsetup} and the corresponding device parameters are presented in Supplementary Table.~\ref{tab1}.

	\subsection{Readout calibration}

	JPA~\cite{Mutus} is used to improve the signal to noise ratio (SNR) of readout signals, and its gain curve is shown in Supplementary Figure~\ref{JPA_gain}.
	The duration time of our readout pulse is 2~$ \mu s $. The transmission data corresponding to five readout resonators are shown, respectively, in the upper row of Supplementary Figure~\ref{readout_calibration}. IQ data at specified frequency values (as indicated by black arrows) are shown in the lower row of Supplementary Figure~\ref{readout_calibration}.  Blue (orange) lines or dots represent data when qubits are prepared at state $|0\rangle$ ($|1\rangle$). There are 2000 dots (repetitions) for each statistic. The readout fidelity $F_{0,j}$ and $F_{1,j}$ of $Q_j$ are listed in Supplementary Table~\ref{tab1}.
	\begin{figure}[!htbp]
		\centering
		\includegraphics[width=0.4\linewidth]{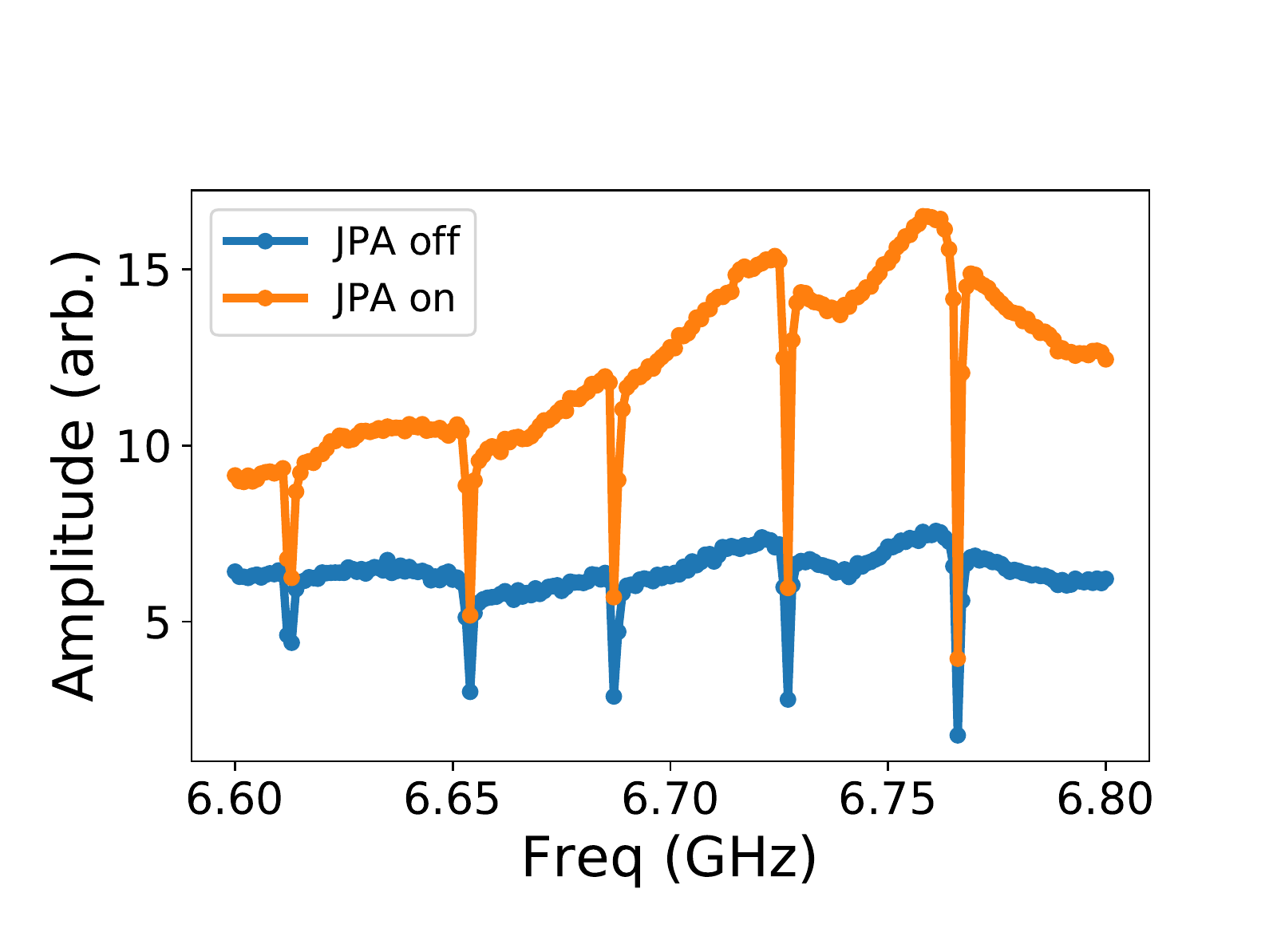}
		\caption{\textbf{Gain of JPA.} ~The blue (orange) line is measured transmission when JPA is turned off (on). Resonant dips correspond to readout resonators respectively. }\label{JPA_gain}		
	\end{figure}

	\begin{figure}[htbp]
		\centering
		\subfigure{
			\begin{minipage}[t]{0.2\linewidth}
				\textbf{$Q_1$}
				\centering
				\includegraphics[width=\linewidth]{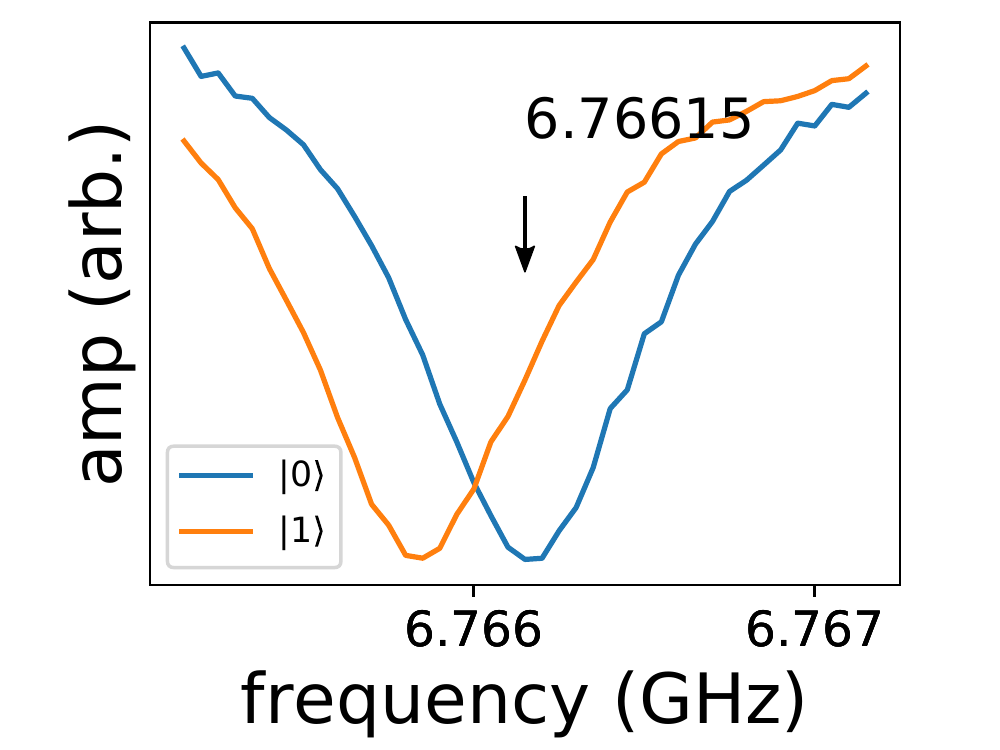}\vspace{4pt}
				\includegraphics[width=\linewidth]{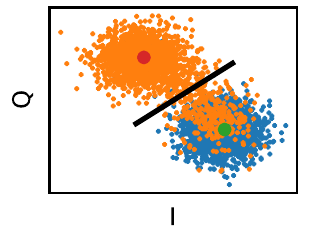}
			\end{minipage}
			\begin{minipage}[t]{0.2\linewidth}
				\textbf{$Q_2$}
				\centering
				\includegraphics[width=\linewidth]{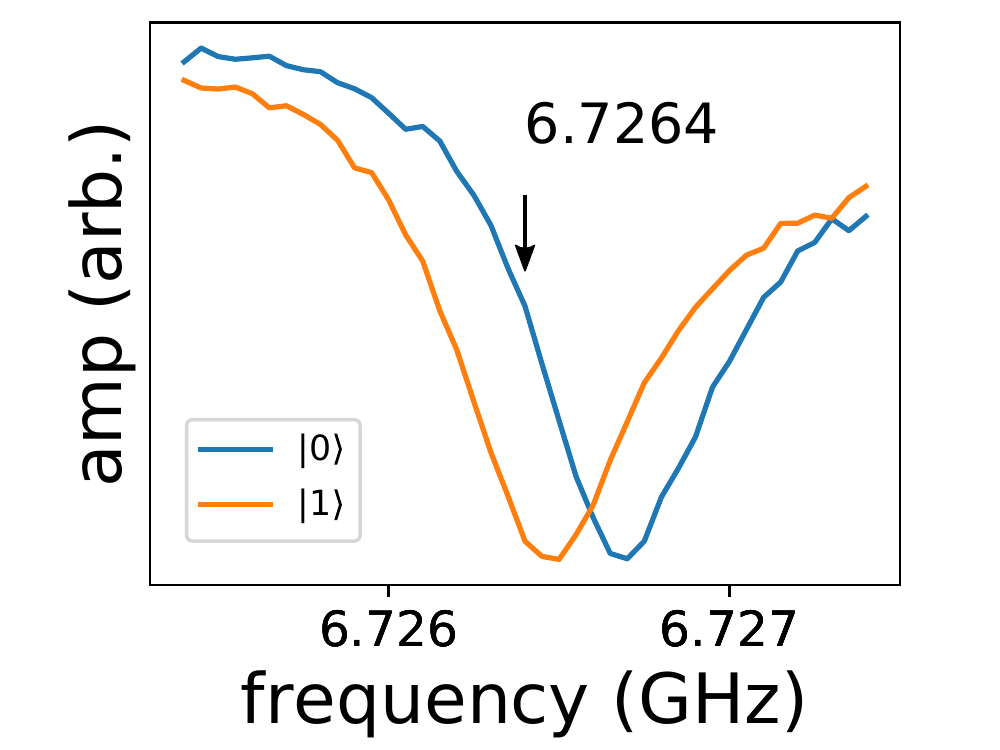}\vspace{4pt}
				\includegraphics[width=\linewidth]{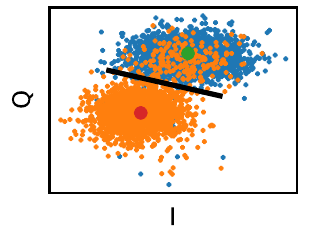}
			\end{minipage}
			\begin{minipage}[t]{0.2\linewidth}
				\textbf{$Q_3$}
				\centering
				\includegraphics[width=\linewidth]{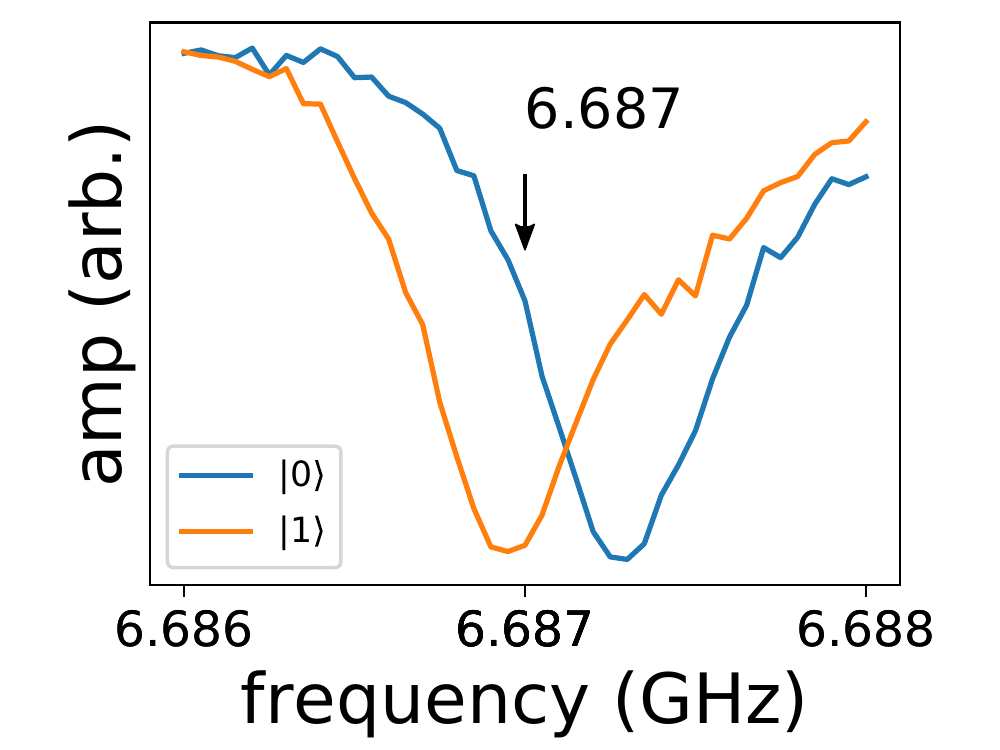}\vspace{4pt}
				\includegraphics[width=\linewidth]{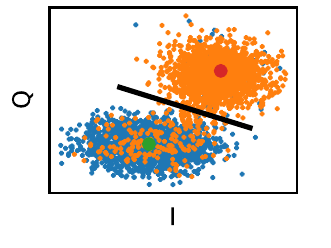}
			\end{minipage}
			\begin{minipage}[t]{0.2\linewidth}
				\textbf{$Q_4$}
				\centering
				\includegraphics[width=\linewidth]{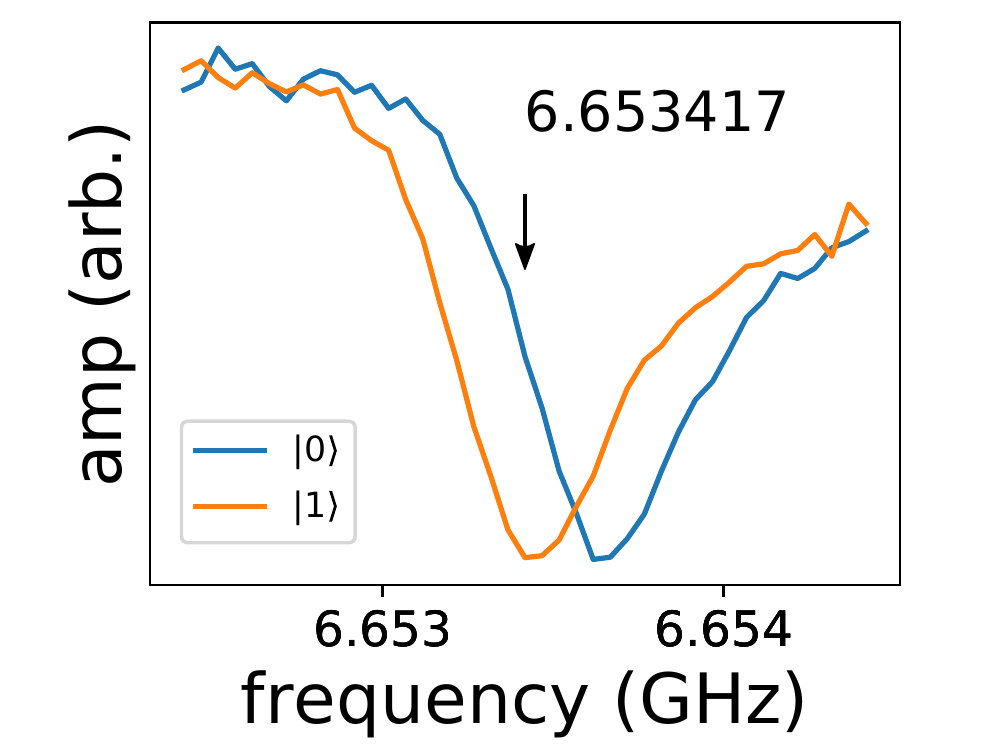}\vspace{4pt}
				\includegraphics[width=\linewidth]{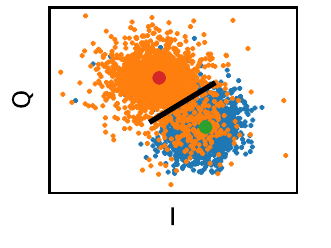}
			\end{minipage}
			\begin{minipage}[t]{0.2\linewidth}
				\textbf{$Q_5$}
				\centering
				\includegraphics[width=\linewidth]{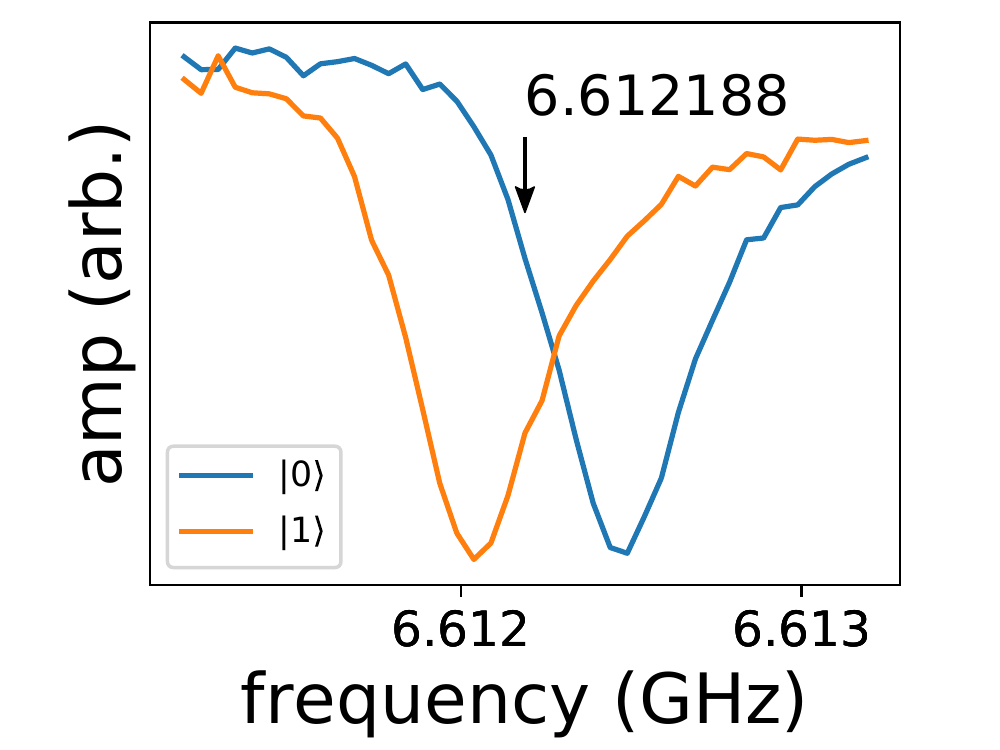}\vspace{4pt}
				\includegraphics[width=\linewidth]{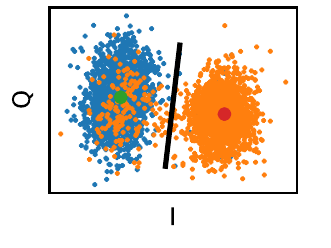}
			\end{minipage}
		}
		\centering
		\caption{\textbf{Qubit readout characteristics.} ~ The transmission data corresponding to five readout resonators are shown, respectively, in the upper row. The IQ data at specified frequency values (as indicated by black arrows) are shown in the lower row. The blue (orange) lines are measured amplitude of readout signals when qubit is prepared in $|0\rangle$ ($|1\rangle$). The complex amplitude of readout signal, is represented by dots in the IQ plane. There are 2000 dots (repetitions) for each statistic. For these characteristics, 5 qubits are measured simultaneously.}\label{readout_calibration}
	\end{figure}

		\newpage
		\mbox{}	

	\subsection{Coherence time}
	Values of energy relaxation time $T_1$ and dephasing time $T_2^*$  for 5 qubits at their idle frequencies are listed in Supplementary Table~\ref{tab1}. The corresponding temporal results are shown in Supplementary Figure~\ref{T1T2}.
	We also measure $ T_1 $ of all 5 qubits around working frequency 4.868 GHz, where the working points of our experiments are in this frequency interval, see Supplementary Figure~\ref{T12d}. The energy relaxation time is long enough to ensure that all 5 qubits have acceptable coherence performance.
	
	\begin{figure}[htbp]
		\centering	
		\subfigure{
			\begin{minipage}[t]{0.2\linewidth}
				\textbf{~~~~~~~~$Q_1$}
				\centering
					
				\text{~~~~~~~$T_1 = 17\mu s$}			
				\includegraphics[width=\linewidth]{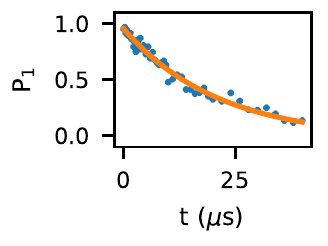}\vspace{4pt}
				\text{~~~~~~~$T_2^* = 1.53\mu s$}
				\includegraphics[width=\linewidth]{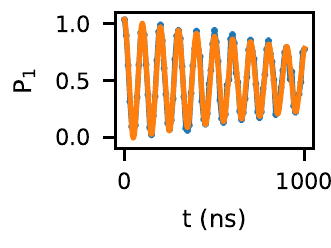}				
			\end{minipage}
			\begin{minipage}[t]{0.2\linewidth}
				\textbf{~~~~~~~~$Q_2$}				
				\centering
				
				\text{~~~~~~~$T_1 = 30\mu s$}
				\includegraphics[width=\linewidth]{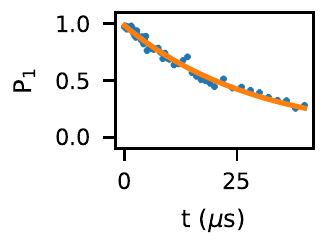}\vspace{4pt}				
				\text{~~~~~~~$T_2^* = 4.39\mu s$}
				\includegraphics[width=\linewidth]{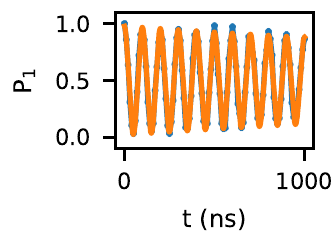}				
			\end{minipage}
			\begin{minipage}[t]{0.2\linewidth}
				\textbf{~~~~~~~~$Q_3$}
				\centering
				
				\text{~~~~~~~$T_1 = 42\mu s$}
				\includegraphics[width=\linewidth]{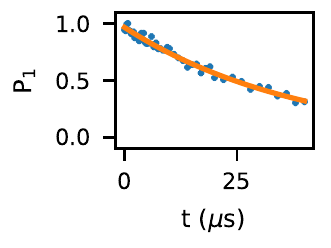}\vspace{4pt}
				\text{~~~~~~~$T_2^* = 2.20\mu s$}
				\includegraphics[width=\linewidth]{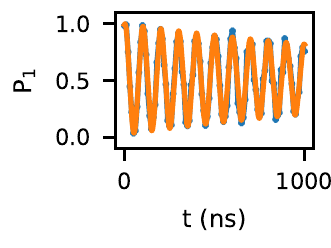}				
				
			\end{minipage}
			\begin{minipage}[t]{0.2\linewidth}
				\textbf{~~~~~~~~$Q_4$}
				\centering
				
				\text{~~~~~~~$T_1 = 17\mu s$}
				\includegraphics[width=\linewidth]{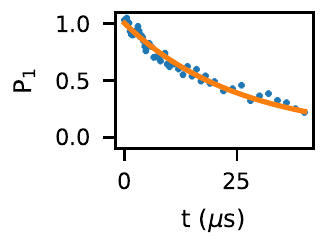}\vspace{4pt}
				\text{~~~~~~~$T_2^* = 2.19\mu s$}
				\includegraphics[width=\linewidth]{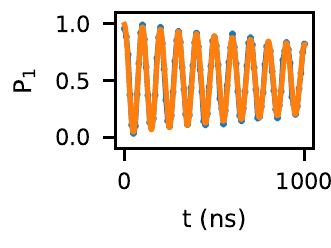}
				
			\end{minipage}
			\begin{minipage}[t]{0.2\linewidth}
				\textbf{~~~~~~~~$Q_5$}
				\centering
				
				\text{~~~~~~~$T_1 = 36\mu s$}
				\includegraphics[width=\linewidth]{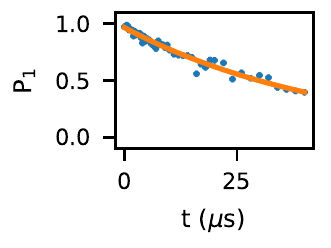}\vspace{4pt}
				\text{~~~~~~~$T_2^* = 2.25\mu s$}
				\includegraphics[width=\linewidth]{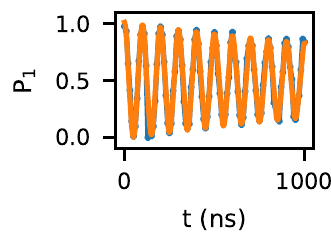}
				
			\end{minipage}
		}				
		\centering
		\caption{\textbf{The results of qubit energy decay and Ramsey interference measurements at idle frequency.} ~Blue dots are measured data and red lines are fitting curve.}	
		\label{T1T2}
	\end{figure}

	\begin{figure}[htbp]
		\centering		
		\subfigure{
			\begin{minipage}[t]{0.2\linewidth}	
				\textbf{$Q_1$}			
				\centering
				\includegraphics[width=\linewidth]{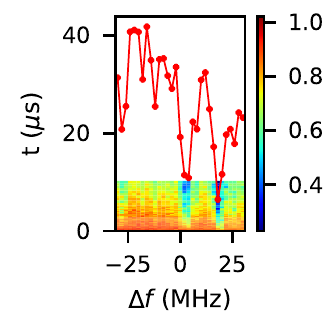}				
			\end{minipage}
			\begin{minipage}[t]{0.2\linewidth}
				\textbf{$Q_2$}
				\centering
				\includegraphics[width=\linewidth]{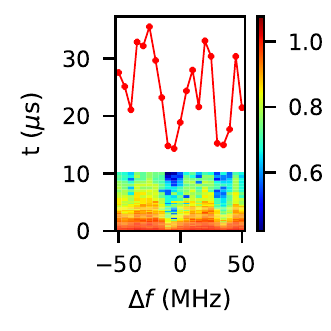}				
	     	\end{minipage}
			\begin{minipage}[t]{0.2\linewidth}
				\textbf{$Q_3$}
				\centering
				\includegraphics[width=\linewidth]{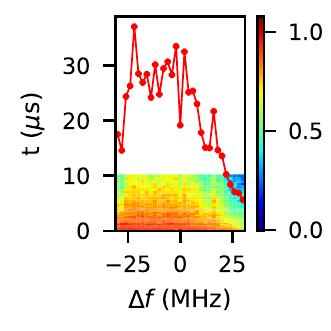}
			\end{minipage}
			\begin{minipage}[t]{0.2\linewidth}
				\textbf{$Q_4$}
	         	\centering
	        	\includegraphics[width=\linewidth]{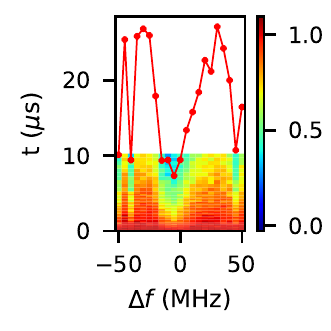}
            \end{minipage}
     		\begin{minipage}[t]{0.2\linewidth}
     			\textbf{$Q_5$}
	     	    \centering
	         	\includegraphics[width=\linewidth]{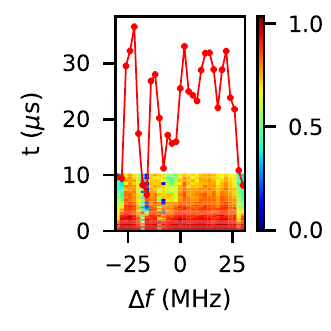}
            \end{minipage}

		}
		\centering
		\caption{\textbf{The experiment results of energy decay measurement around working frequency.} ~$\Delta f$ is the frequency difference to 4.868 GHz. The red dots are $T_1$ values, evaluated from exponential decay fitting of the results. }	\label{T12d}		
	\end{figure}

	\subsection{Z control line crosstalk}
	
    Each qubit is tuned by current pulses applied to its Z control line. However, there are crosstalks between Z control lines, which means that the current pulse on one Z control line can also affect other qubits. To evaluate the influence of such crosstalk, we bias target qubit $Q_i$ at $f_i$ which is more than 1 GHz below its maximum frequency, so that the qubit frequency is sensitive to external magnetic flux  as well as the Z pulse crosstalk. We apply a long ($2 \mu s$) Gaussian shaped resonant X ($\pi$) pulse, and then measure the excitation probability of $Q_i$. Applying a Z pulse with amplitude $Z_j$ to the Z control line of another qubit $Q_j$ will generate a crosstalk pulse on $Q_i$, as shown in Supplementary Figure~\ref{z_compensation}. Such a crosstalk pulse would lead to a frequency change for $Q_i$, and hence the decrease of the excitation probability. We apply a compensation pulse with amplitude $Z^{comp}_1$  to the Z control line of $Q_i$ to cancel this crosstalk. In practice, we monitor the excitation probability as a function of compensation pulse amplitude. When a complete compensation is achieved, the excitation probability is maximized.  The linear relation  between  $Z_j$ and  {$Z^{comp}_1$ is shown in Supplementary Figure~\ref{z_compensation_result}.

	\begin{figure}[!htp]
		\centering
		\includegraphics[width=0.65\linewidth]{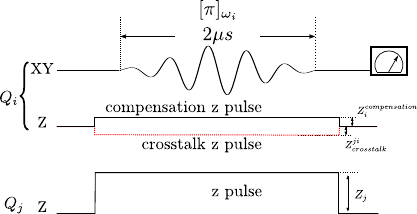}
		\caption{\textbf{Schematic diagram for Z control line crosstalk compensation.} ~  $ Q_i $ is bias to a magnetic flux sensitive point with a frequency $ f_i $. When a square pulse with amplitude of $ Z_j $ is applied to another qubit $ Q_j $,  it could generate a crosstalk pulse with amplitude  $ Z ^ {ji} _ {crosstalk} $. To cancel out the influence caused by this crosstalk and to maintain the qubit frequency to $ f_i $, we apply $ Z^{comp}_i = -Z ^ {ji} _ {crosstalk} $ amplitude to the Z control line for $ Q_i $. }\label{z_compensation}
	\end{figure}

	\begin{figure}[htbp]
		\centering
		\subfigure[]{		
		\begin{minipage}[t]{0.25\linewidth}
			
				\centering
				\includegraphics[width=\linewidth]{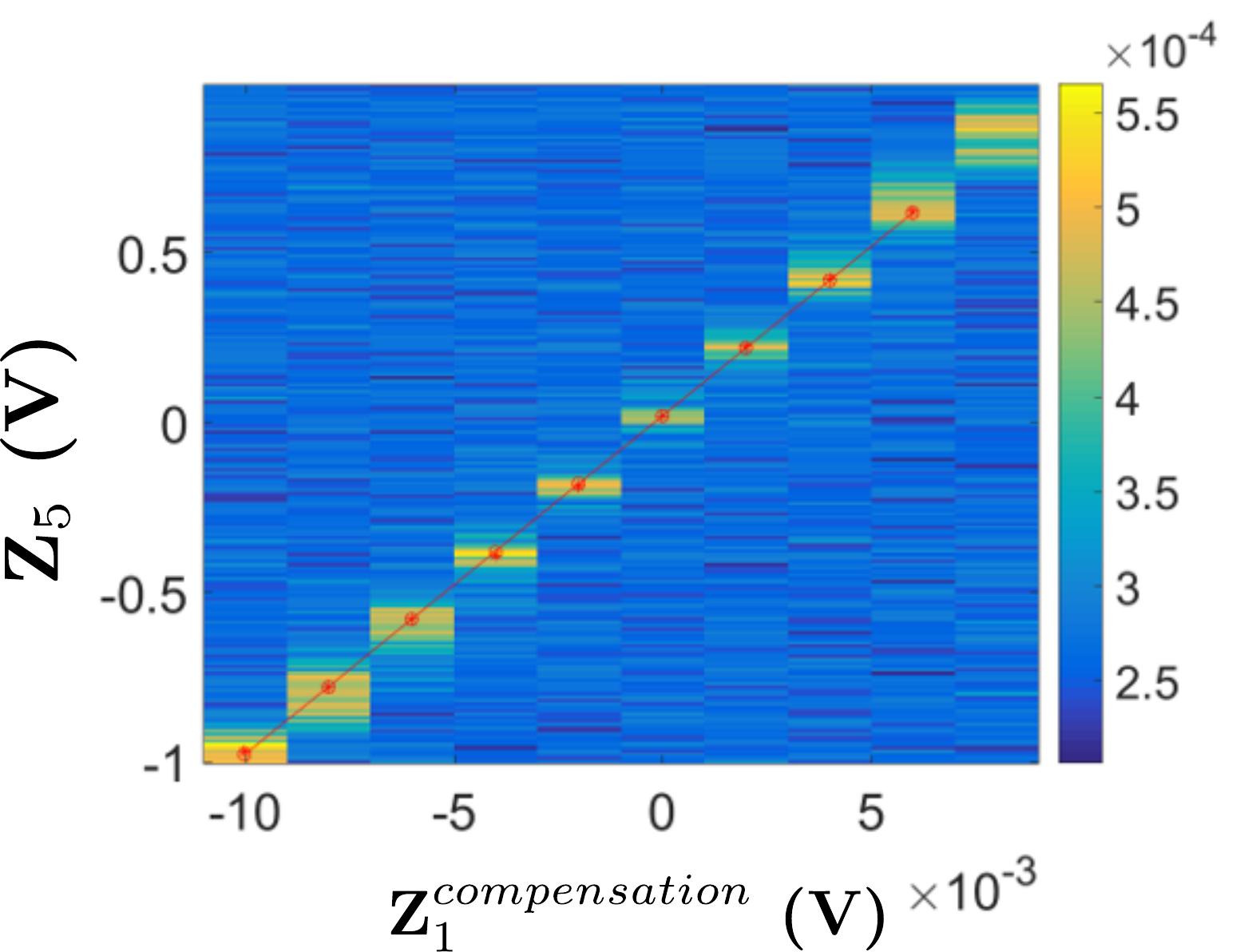}

		\end{minipage}	
		\begin{minipage}[t]{0.25\linewidth}

				\centering
				\includegraphics[width=\linewidth]{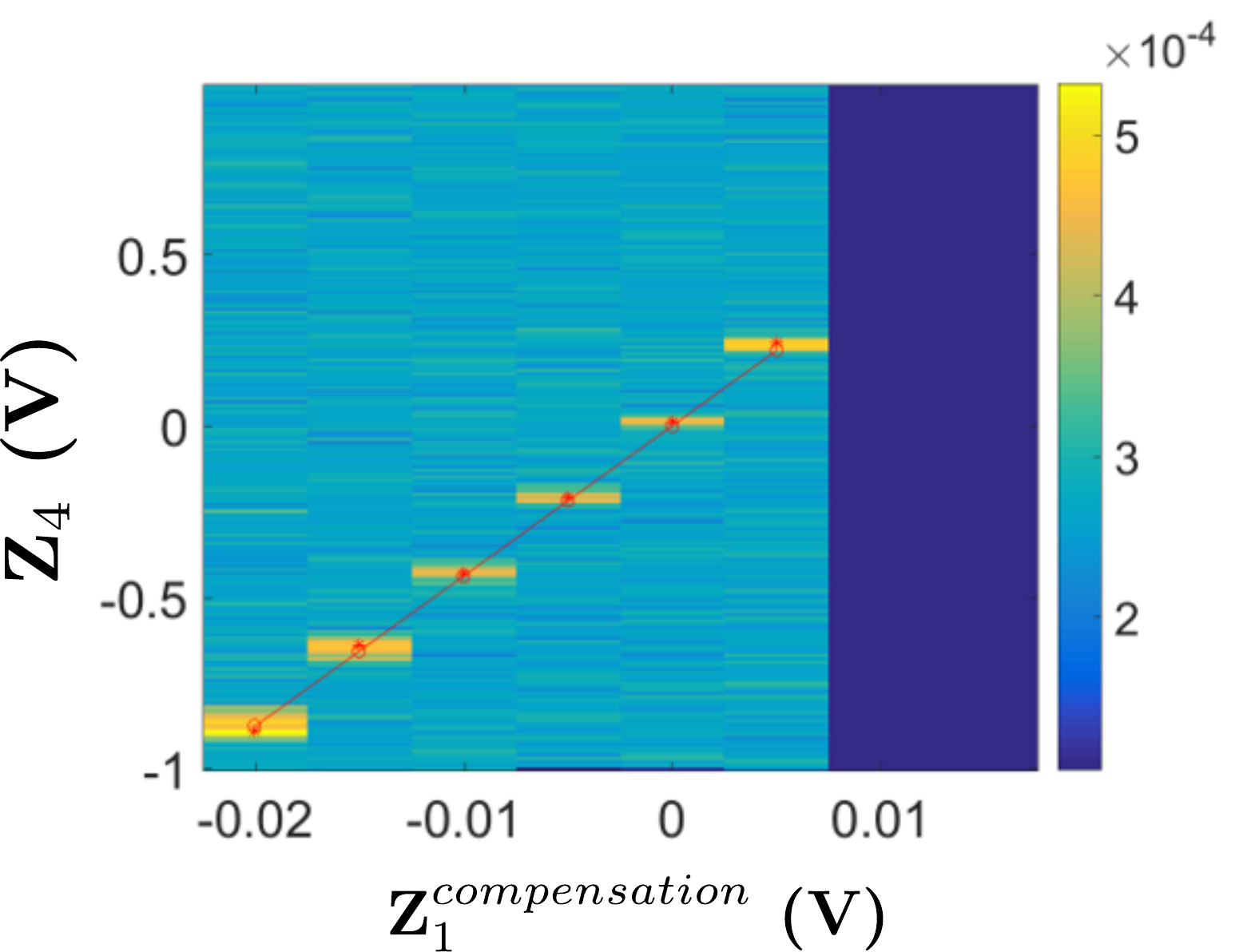}

		\end{minipage}	
		\begin{minipage}[t]{0.25\linewidth}

				\centering
				\includegraphics[width=\linewidth]{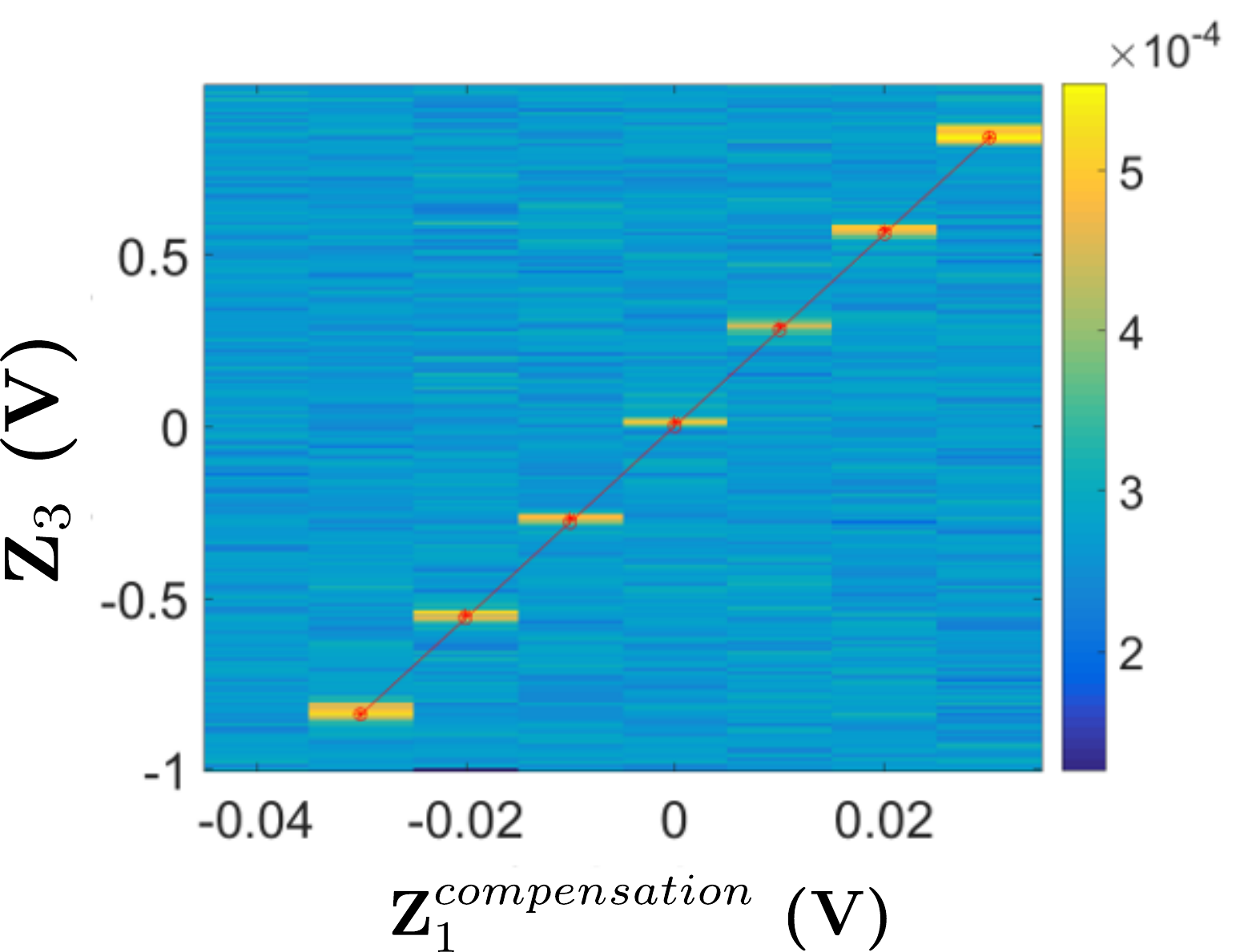}

        \end{minipage}
		\begin{minipage}[t]{0.25\linewidth}

				\centering
				\includegraphics[width=\linewidth]{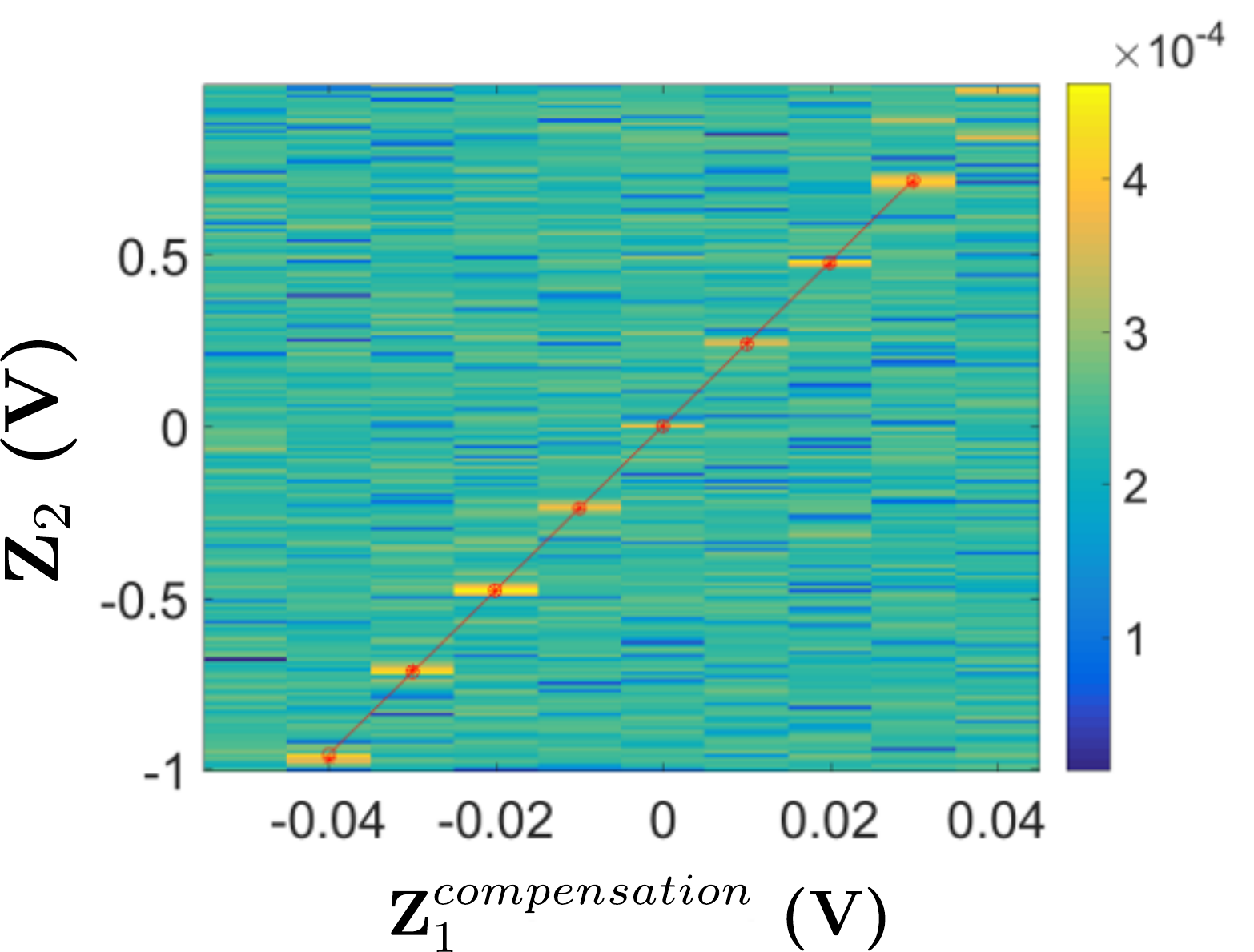}

		\end{minipage}
	}
		\centering
		\caption{\textbf{Results for determining crosstalk-compensation coefficient of $Q_1$.} ~Changing $Z_j (j\neq 1)$ and $Z^{comp}_1$ to measure the excitation probability of $Q_1$. The compensation coefficients are obtained by fitting the linearly varied data points. } \label{z_compensation_result}	
	\end{figure}
	
    After completing such crosstalk-compensation measurement for all qubits, a full transformation matrix for Z control line crosstalk calibration is constructed and shown in Eq.\ref{z_compensation_matrix}, where $ Z_i $ represents the actual amplitude applied on the Z control line of $ Q_i $, and $ Z_i ^ * $ represents the effective amplitude after taking crosstalk compensation into account.	
	
	\begin{equation}
	\begin{bmatrix}
	z_1^*\\
	z_2^*\\
	z_3^*\\
	z_4^*\\
	z_5^*
	\end{bmatrix}
	=
	\begin{bmatrix}
	1.000 & -0.042 & -0.036 & -0.022 & -0.010 \\
	-0.065 & 1.000 & 0.064 & 0.065 & 0.061 \\
	-0.054 & -0.107 & 1.000 & 0.140 & 0.097 \\
	0.041 & 0.067 & 0.113 & 1.000 & -0.119 \\
	-0.015 & -0.025 & -0.037 & -0.020 & 1.000
	\end{bmatrix}
	\cdot
	\begin{bmatrix}
	z_1\\
	z_2\\
	z_3\\
	z_4\\
	z_5
	\end{bmatrix}\label{z_compensation_matrix}
	\end{equation}

	\subsection{Calibration of time delay for all control channels}
	
	To guarantee the pulses applied on different control channels reaching to chip at the expected time, the output delay time of AWG ports need to be adjusted to compensate the transmission time difference in different channels. In our experiment, we set Z control channel of $Q_1$ as the reference, and make all other control channels align to it.
	There are two steps:
	In the first step, we align Z control channels by aligning $Q_2$ to $Q_1$, followed by $Q_3$ to $Q_2$ and so forth. In the second step, we align XY control channel of each qubit to  its   Z control channel.
	
	Scheme for determining Z control channel delay length is shown in Supplementary Figure~\ref{Z_control_line_delay}.
	The blue lines represent transmission time in control channels. The red lines are the output delays that we insert before AWG output. The black dots are the AWG output start times. By adjusting the length of output delay, we can make the Z pulses align to the reference channel. In experiment, we set $Q_{i}$ (or $Q_{i+1}$) to $|1\rangle$, then apply two square pulses to make $Q_i$ and $Q_{i+1}$  resonantly coupled  for a time duration $2\pi/4g$. This means  that if two pulses are aligned, the resonant coupling will induce a complete photon swap. Here $g$ is the coupling strength between $Q_i$ and $Q_{i+1}$. During the measurement, we vary the length of output delay time of $Q_{i+1}$ and measure $Q_i$ (or $Q_{i+1}$) after the Z pulses. The final state of $Q_i$ (or $Q_{i+1}$) depends on the overlap of two Z pulses. When two pulses are aligned, the overlap will be equal to  $2\pi/4g$, and the swap probability will be the highest.
	The experimental results and fittings are shown in Supplementary Figure~\ref{Z_control_line_delay_result}. In our experimental setup, the cables connecting qubits and electronics are designed to be  {\color{red} of}  the same length, so the final adjustment of the output delay is in the range of  $\pm500 $ps.

	\begin{figure}[!htp]
		\centering
		\includegraphics[width=0.65\linewidth]{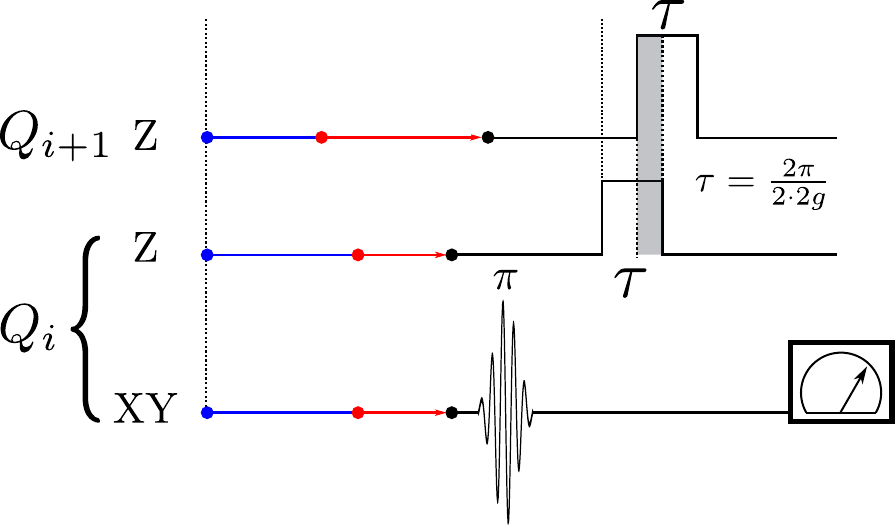}
		\caption{\textbf{Schematic diagram for arriving time of pulses to the qubit.} ~ Blue lines represent the transmission times of control channels. Red lines represent the individually adjustable delays inserted before AWG output. Black dots represent the output start time of AWG ports.  By adjusting the length of delay, we can align the square pulse on $Q_{i+1}$ Z control channel to square pulse on $Q_i$ Z control channel. The  shadow part represents the time overlap of two pulses.}
		\label{Z_control_line_delay}
	\end{figure}
	
	After the alignment of all Z control channels, we align XY control channel of each qubit to its Z control channel in time. As shown in Supplementary Figure~\ref{XY_Z_t0}, the effective excitation probability of $Q_i$ is the integral of Gaussian shaped $\pi$ pulse over time before the Z pulse. When we adjust the output delay of XY channel, and measure the final state of $Q_i$, we obtain the data in Supplementary Figure~\ref{XY_Z_t0_result}. By fitting the data, we can get the proper output delay time of XY channel. With cables of the same length, the measured delay times are also small and in the range of $\pm500 $ps.

	\begin{figure}[!htp]
		\centering		
		\includegraphics[width=0.65\linewidth]{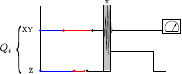}
		\caption{\textbf{Schematic diagram for aligning XY channel to Z channel.} ~ The same as Fig.~\ref{Z_control_line_delay}, we use blue lines to represent the transmission time through control channels, and use red lines to represent the individually adjustable delay before AWG output ports. By adjusting the length of delay, we can align $\pi$ pulse in XY channel to square pulse in Z channel.
		}
		\label{XY_Z_t0}
	\end{figure}

	\begin{figure}[htbp]
		\centering				
		\subfigure{
			\begin{minipage}[t]{0.225\linewidth}
				\textbf{~~~~~~$Q_1-Q_2$}
				\centering
				\includegraphics[width=\linewidth]{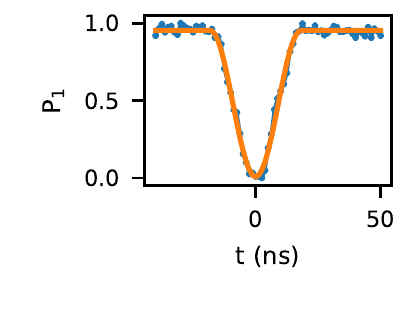}
			\end{minipage}
			\begin{minipage}[t]{0.225\linewidth}
				\textbf{~~~~~~$Q_2-Q_3$}
				\centering
				\includegraphics[width=\linewidth]{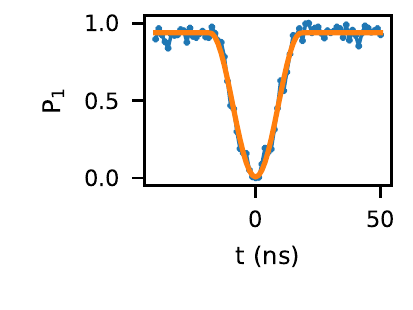}
		   \end{minipage}
			\begin{minipage}[t]{0.225\linewidth}
				\textbf{~~~~~~$Q_3-Q_4$}
				\centering
			    \includegraphics[width=\linewidth]{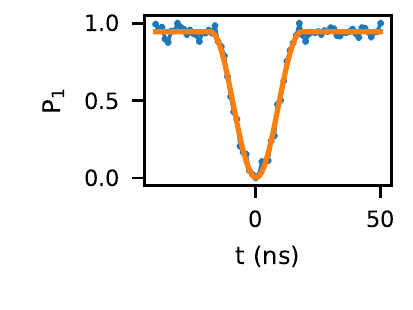}
	        \end{minipage}	
			\begin{minipage}[t]{0.225\linewidth}
				\textbf{~~~~~~$Q_4-Q_5$}
				\centering
				\includegraphics[width=\linewidth]{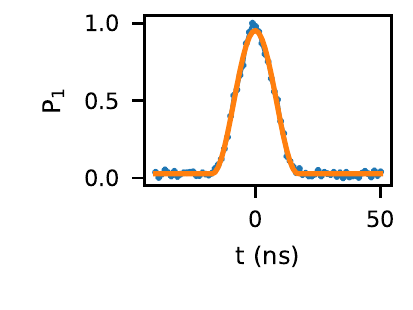}
			\end{minipage}

		}						
		\centering
		\caption{\textbf{Results of Z channels alignment experiments in Fig.~\ref{Z_control_line_delay}.} }\label{Z_control_line_delay_result}				
	\end{figure}

	\begin{figure}[htbp]
		\centering				
		\subfigure{
			\begin{minipage}[t]{0.2\linewidth}
				\textbf{~~~~~~$Q_1$}
	         	\centering
	        	\includegraphics[width=\linewidth]{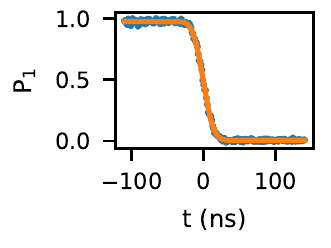}
            \end{minipage}
			\begin{minipage}[t]{0.2\linewidth}
				\textbf{~~~~~~$Q_2$}
				\centering
				\includegraphics[width=\linewidth]{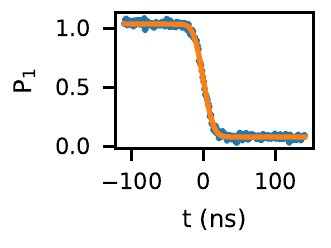}
            \end{minipage}
			\begin{minipage}[t]{0.2\linewidth}
				\textbf{~~~~~~$Q_3$}
				\centering
				\includegraphics[width=\linewidth]{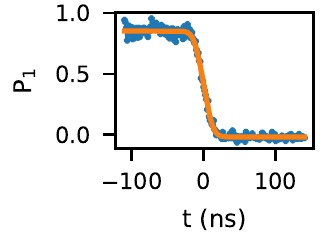}
			\end{minipage}
			\begin{minipage}[t]{0.2\linewidth}
				\textbf{~~~~~~$Q_4$}
				\centering
				\includegraphics[width=\linewidth]{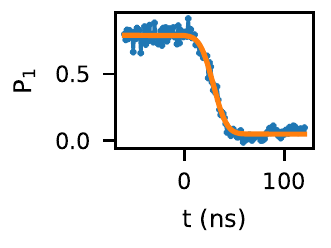}
              \end{minipage}
			\begin{minipage}[t]{0.2\linewidth}
				\textbf{~~~~~~$Q_5$}
				\centering
				\includegraphics[width=\linewidth]{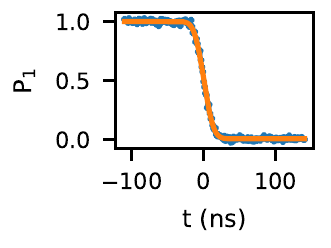}
			\end{minipage}
		}		
		\centering
		\caption{\textbf{Results of XY channel to Z channel alignment experiments in Fig.~\ref{XY_Z_t0}.} }\label{XY_Z_t0_result}					
	\end{figure}
	
%	
%	\newpage
%	\mbox{}		
	\subsection{Square pulse distortion correction}
	
	In our experiment, qubits are tuned to specified frequencies by square pulses. However, an ideal square pulse is usually distorted when reaching to the chip. To correct this distortion, we need to determine the deformed shape of the step edge of Z square pulse. The corresponding qubit is biased to a relative low frequency $f_z$ (more than 1GHz below its maximum frequency) to improve its sensitivity to the pulse shape deformation. As shown in Supplementary Figure~\ref{Z_pulse_shape}, an amplitude fixed ($Z_{amp}$) step signal is applied to a qubit, and followed a $\pi$ pulse with frequency $f_z$ and length 20ns. If there is no distortion, the qubit will be excited to $|1\rangle$ by the $\pi$ pulse. If there is small distortion after the step signal, qubit will not be completely excited to $|1\rangle$. By varying the compensation amplitude $\Delta Z$, we can find the value of full compensation when the maximum excitation probability is achieved. By changing the delay time $t$ of $\pi$ pulse, we can get a view of the distorted step signal response, as shown in the upper row of Supplementary Figure~\ref{z_pulse_data}.  With this response data, we can calculate the needed adjustments for the square pulses. Then, we repeat the measurement in Supplementary Figure~\ref{Z_pulse_shape} with corrected pulses , and find the step signal response is flattened, see the lower row of Supplementary Figure~\ref{z_pulse_data}.

	\begin{figure}[!htp]
		\centering
		\includegraphics[width=0.65\linewidth]{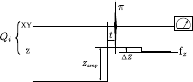}
		\caption{\textbf{Schematic diagram for measuring the step signal response.} ~Qubit is biased to a low frequency $f_z$ (usually 1 GHz lower than its maximum frequency) to gain better sensitivity to flux variation.  $Z_{amp}$ is fixed. The length of Gaussian shaped $\pi$ pulse is set to 20ns for acceptable time resolution. The frequency of $\pi$ pulse is set to $f_z$ and fixed. In experiment, for different delay time $t$, $\Delta Z$ are varied to find the full compensation to the distortion. When qubit excitation probability reaches the maximum, full compensation is achieved. }\label{Z_pulse_shape}
	\end{figure}
	
	\begin{figure}[htbp]
		\centering
		\subfigure{
			\begin{minipage}[t]{0.2\linewidth}
				\textbf{$Q_1$}
				\centering
				\includegraphics[width=\linewidth]{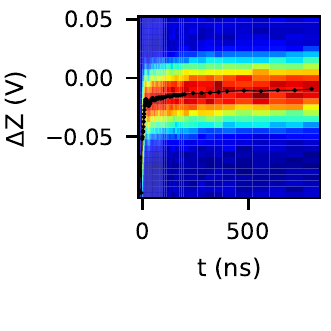}\vspace{4pt}
				\includegraphics[width=\linewidth]{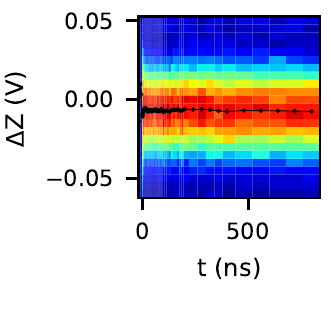}
			\end{minipage}
			\begin{minipage}[t]{0.2\linewidth}
				\textbf{$Q_2$}
			\centering
			\includegraphics[width=\linewidth]{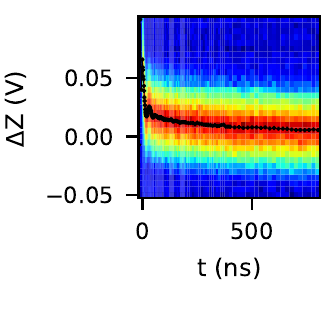}\vspace{4pt}
			\includegraphics[width=\linewidth]{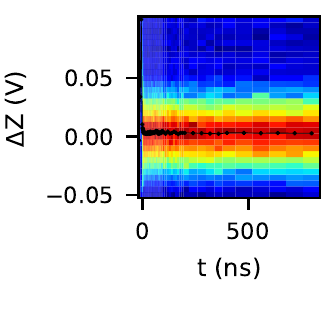}
		    \end{minipage}
			\begin{minipage}[t]{0.2\linewidth}
				\textbf{$Q_3$}
				\centering
				\includegraphics[width=\linewidth]{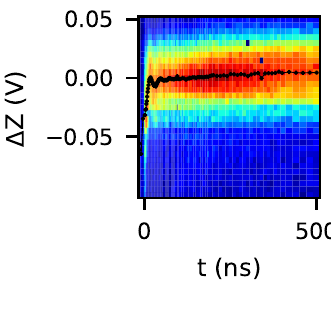}\vspace{4pt}
				\includegraphics[width=\linewidth]{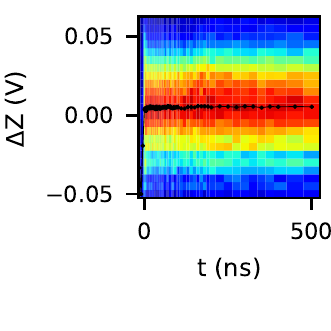}			
			\end{minipage}
			\begin{minipage}[t]{0.2\linewidth}
				\textbf{$Q_4$}
	     	\centering
	    	\includegraphics[width=\linewidth]{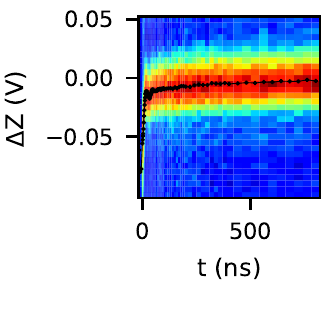}\vspace{4pt}
	      	\includegraphics[width=\linewidth]{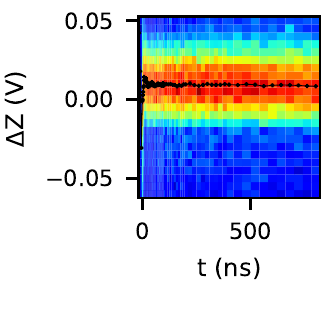}
      	   \end{minipage}
           \begin{minipage}[t]{0.2\linewidth}
           	\textbf{$Q_5$}
         	\centering
         	\includegraphics[width=\linewidth]{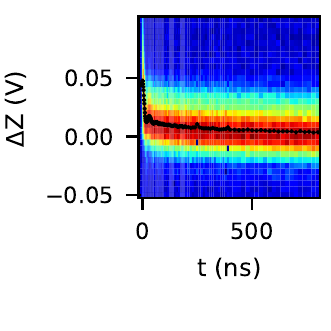}\vspace{4pt}
         	\includegraphics[width=\linewidth]{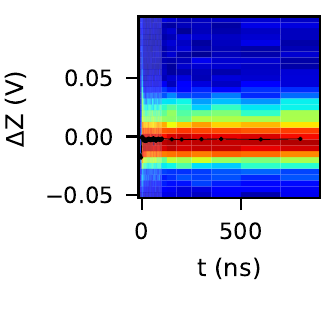}
           \end{minipage}
		}					
		\centering
		\caption{\textbf{Results of step signal response measurement in Fig.~\ref{Z_pulse_shape}.} ~ The upper row are results before correction. Black dots, corresponding to maximum excitation, show the shape of step edge. The lower row are results after correction.}	
		\label{z_pulse_data}
	\end{figure}

	\subsection{Calibration of initial state phase}
	
	In order to measure the energy density transport, we need to prepare 5 qubits in an initial state such as $ |X_+X_+ 000 \rangle $, and measure the expectation value of correlated operator $\sigma^i_x\sigma^{i+1}_x$ ($\sigma^i_y\sigma^{i+1}_y$). In the experiment, we firstly rotate $Q_1$ and $Q_2$ to X-Y plane with a $\pi/2$ pulse and set the initial state of $Q_1$ as $\ket{X_+}$,  then we need to adjust the state of $Q_2$ by changing the phase $\phi$ of $\pi/2$ pulse to make $Q_2$ at $\ket{X_+}$, see Supplementary Figure~\ref{InitialPhaseCalibration}. Here, we use the Bloch sphere representation for describing the state of a qubit.  Then these two qubits are resonantly coupled to each other at the frequency $f_c$. (In our experiment, we need to resonantly couple all 5 qubits at this frequency, and set a linear frequency gradient around $f_c$.) \space After time $\tau = 2\pi/4\cdot2g$, we measure the final state probability  as a function of  the phase $\phi$. The corresponding result is shown in the leftmost figure in Supplementary Figure~\ref{InitialPhaseCalibrationResult}. The curves of $P_{|01\rangle}$ and $P_{|10\rangle}$ are consistent with  theoretical and numerical analysis.
The cross point, indicated by the black arrow, corresponds to the $\phi$ when the state  of $Q_2$ is also $\ket{X_+}$.
	In a similar way, we can calibrate the phase between $Q_3$ and $Q_2$, and so forth.

	\begin{figure}[!htp]
		\centering
		\includegraphics[width=0.6\linewidth]{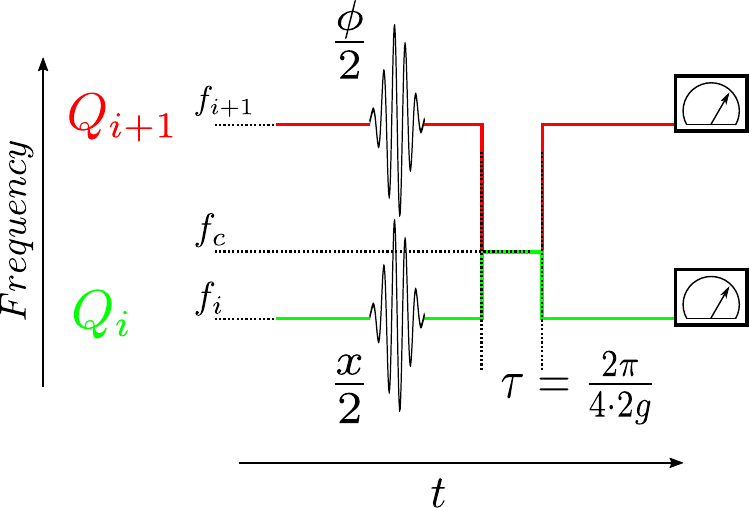}
		\caption{ \textbf{Schematic diagram for calibrating initial state phases of qubits.} ~Rotate both qubits to X-Y plane, and set $Q_i$ initial direction as the reference $X_+$. Adjust the phase (direction) of $Q_{i+1}$, and make them resonantly couple for a time duration $\tau = 2\pi/4\cdot2g$,  where $g$ is the coupling strength between two qubits. At the end, both qubits are measured.}\label{InitialPhaseCalibration}
	\end{figure}

	\begin{figure}[htbp]
		\centering
		\subfigure{
			\begin{minipage}[t]{0.25\linewidth}
				\textbf{~~~~~~$Q_1-Q_2$}
				\centering
				\includegraphics[width=\linewidth]{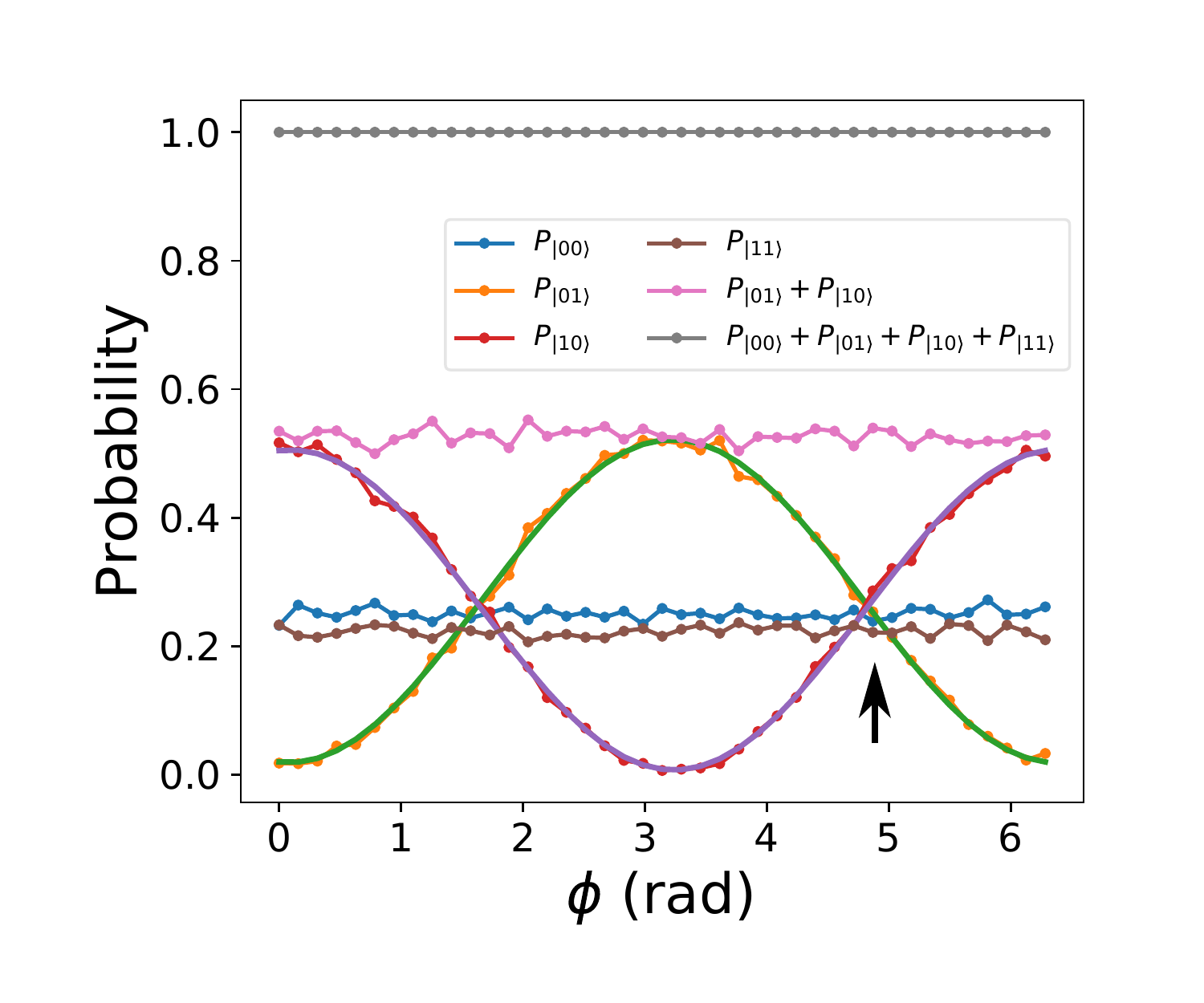}
			\end{minipage}
			\begin{minipage}[t]{0.25\linewidth}
				\textbf{~~~~~~$Q_2-Q_3$}
				\centering
				\includegraphics[width=\linewidth]{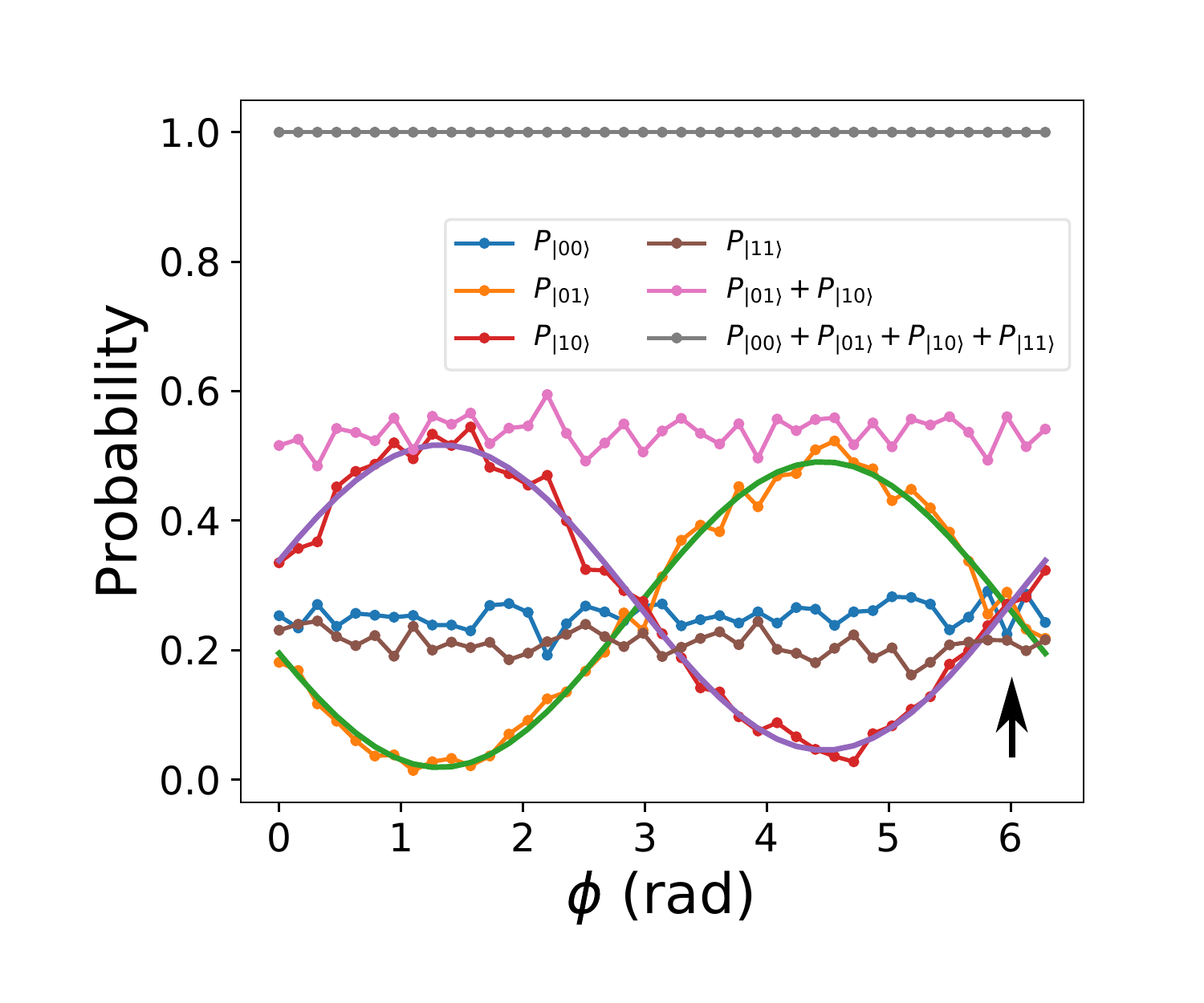}
			\end{minipage}
			\begin{minipage}[t]{0.25\linewidth}
				\textbf{~~~~~~$Q_3-Q_4$}
				\centering
				\includegraphics[width=\linewidth]{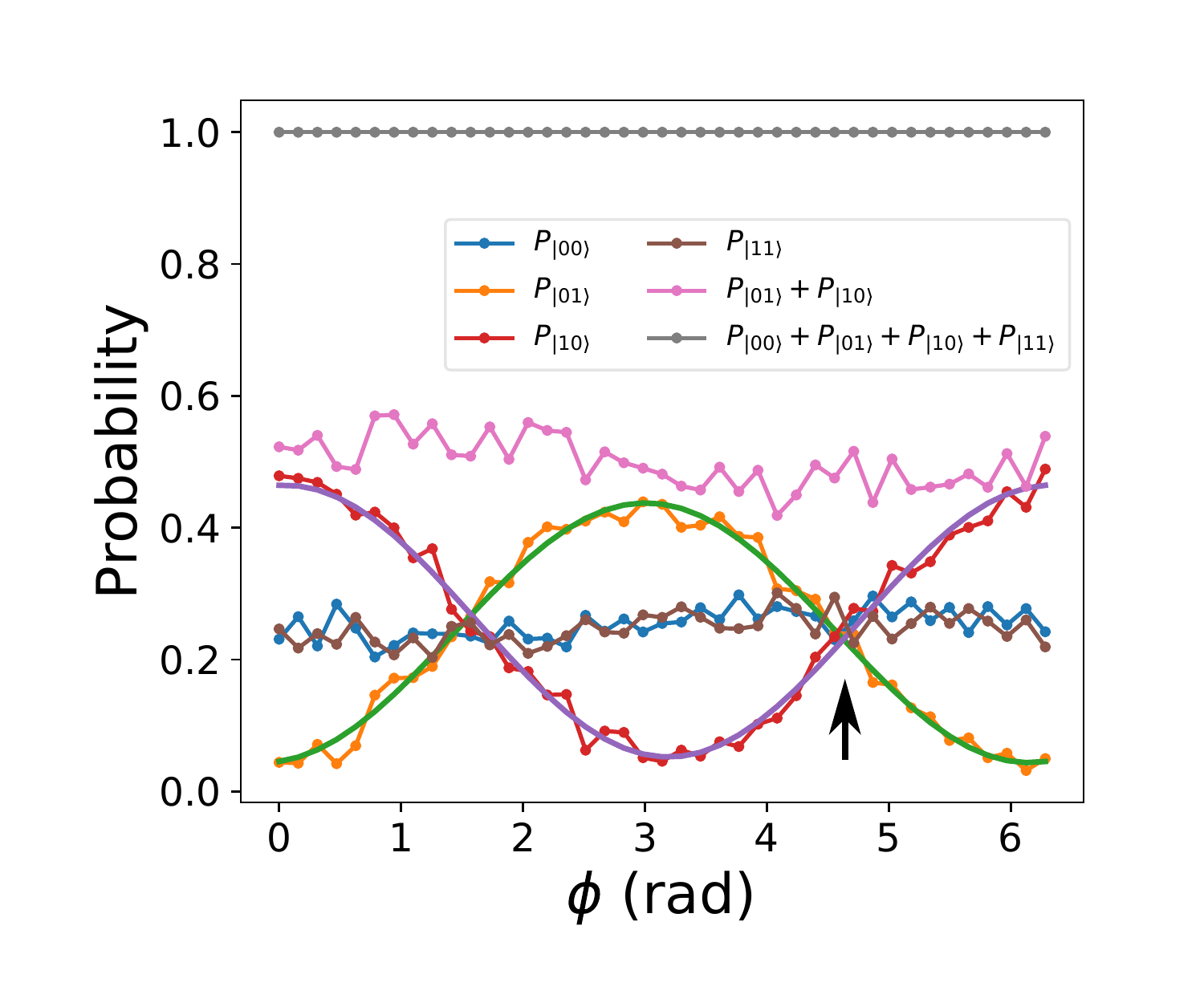}
			\end{minipage}
			\begin{minipage}[t]{0.25\linewidth}
				\textbf{~~~~~~$Q_4-Q_5$}
				\centering
				\includegraphics[width=\linewidth]{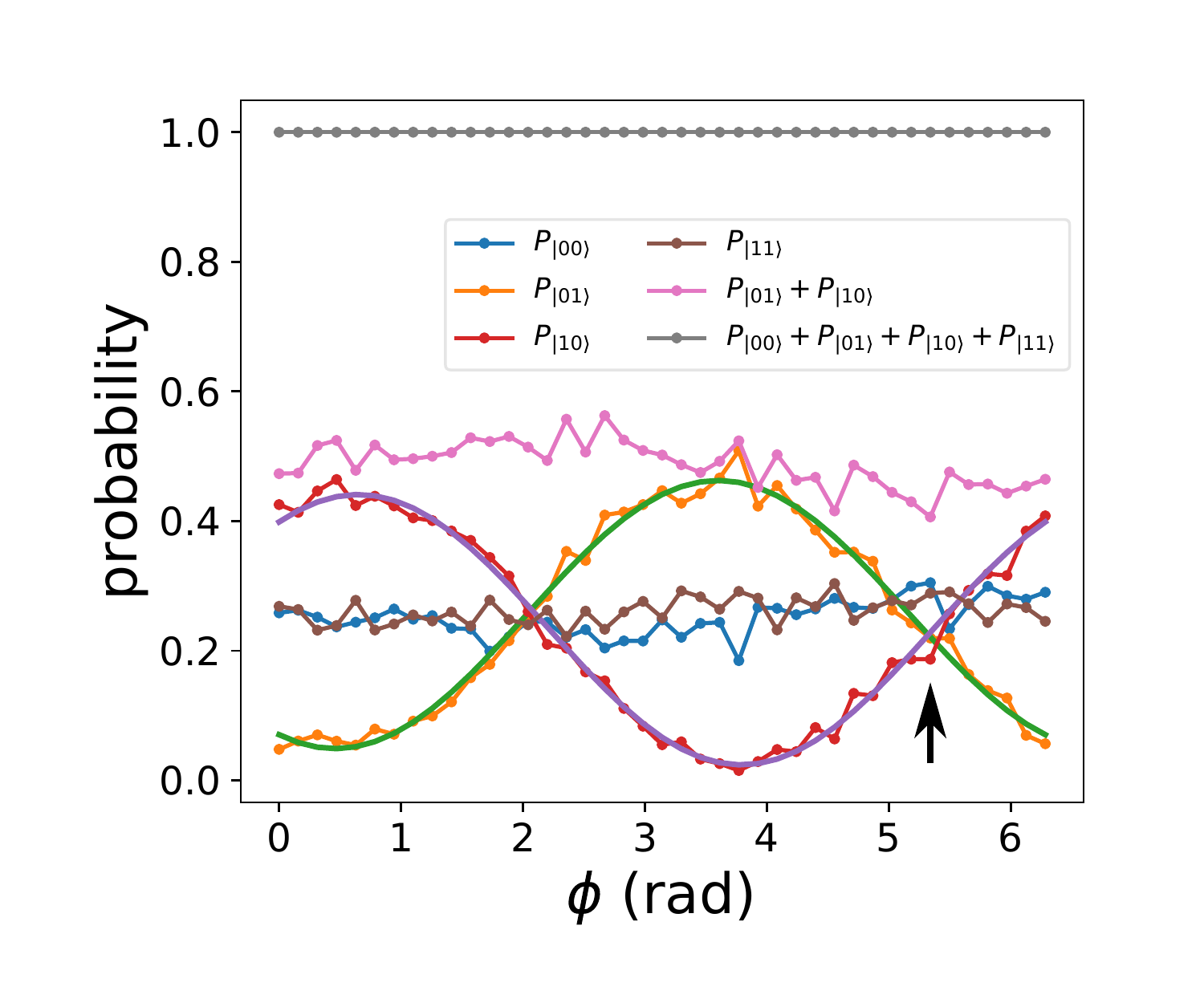}
			\end{minipage}
		}				
		\centering
		\caption{\textbf{Results of experiment in Fig.~\ref{InitialPhaseCalibration}.}  ~For $Q_i$-$Q_{i+1}$, the phase of $Q_i$ is fixed and set as reference $X_+$. Varying the phase $\phi$ of $Q_{i+1}$ from 0 to $2\pi$, the probability distribution of final states varies as a function of $phi$. The black arrows show the phase values for qubits when they are in $X_+$ direction. }\label{InitialPhaseCalibrationResult}
	\end{figure}
	
	\subsection{Calibration of dynamical phase}
	
%	We have calibrated the phase of initial state $ |X_+X_+ 000 \rangle $.
	To measure the expectation value of correlated operator $\sigma^i_x\sigma^{i+1}_x$ ($\sigma^i_y\sigma^{i+1}_y$), we need to determine the phase of final state for each qubit. In our experiment, qubits are coupled at frequency $f_c$,  but are rotated and measured at their individual idle frequencies $f^{idle}$.  The frequency difference between $f_c$ and $f^{idle}$ can lead to   a dynamical phase accumulation to the final state. Thus, in our experiment, we need to calibrate this accumulated dynamical phase.
	
	The accumulation of the dynamical phase can be divided into three parts: rising edge part, middle flat part, and falling edge part. The summation, represented by the shadowed part in Supplementary Figure~\ref{DyPhaseDetermination}, can be expressed as $ \phi (\tau) = \int_ {0} ^ {\tau_ {r}} \Delta \omega_ {r} (t) dt + \Delta \omega \cdot (\tau-\tau_ {r} -\tau_ {f}) + \int _ {\tau-\tau_ {f}} ^ {\tau} \Delta \omega_ {f } (t) dt $. Due to the presence of the rising and falling edges,  the actual dynamical phase accumulation, comparing with   $ \Delta \omega \cdot \tau $, has an approximately fixed offset $ \phi_d = \phi (\tau)- \Delta \omega \cdot \tau $. To determine $ \phi(\tau) $, we use the Ramsey-like method as shown in Supplementary Figure~\ref {DyPhaseDetermination}. The phase of the first $ \pi / 2 $ pulse is $\phi_0 $ and fixed, and the phase of the second $ \pi / 2 $ pulse is $ \phi_0  + \Delta \omega \cdot \tau + \varphi$. Thus, when $ \varphi = \phi_d $, the phase of the second $ \pi / 2 $ pulse is equivalent to $ \phi_0 + \phi(\tau)$, and the qubit will be in $|1\rangle$.
	
	The calibrated results of $ Q_1 $, $ Q_2 $, $ Q_4 $, and $ Q_5 $ are shown in Supplementary Figure~\ref{DyPhaseDeterminationResult}.
	
	\begin{figure}[!htp]
		\centering
		\includegraphics[width=0.6\linewidth]{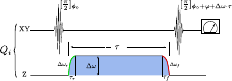}
		\caption{\textbf{Ramsey-like experiment for determining the dynamical phase accumulation.} ~$\Delta\omega_r$ ($\Delta\omega_f$) corresponds to the rising (falling) edge parts of the square pulse. 	To illustrate clearly, raising and falling edge parts are exaggerated in the diagram.
		 $\Delta\omega_r$, $\tau_r$, $\Delta\omega_f$, and $\tau_f$ are approximately unchanged when $\tau$ is increased. The area of the square pulse is $ \phi (\tau) = \int_ {0} ^ {\tau_ {r}} \Delta \omega_ {r} (t) dt + \Delta \omega \cdot (\tau-\tau_ {r} -\tau_ {f}) + \int _ {\tau-\tau_ {f}} ^ {\tau} \Delta \omega_ {f } (t) dt $.}\label{DyPhaseDetermination}
	\end{figure}

	\begin{figure}[htbp!]	
		\centering	
		\subfigure{
			\begin{minipage}[t]{0.25\linewidth}
				\textbf{$Q_1$}
				\centering
				\includegraphics[width=\linewidth]{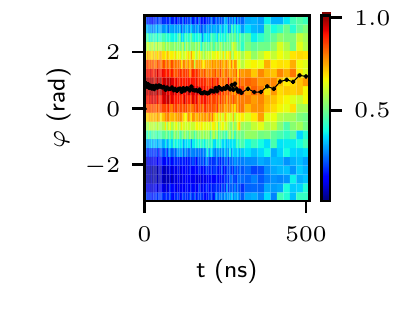}
			\end{minipage}
			\begin{minipage}[t]{0.25\linewidth}
				\textbf{$Q_2$}
				\centering
				\includegraphics[width=\linewidth]{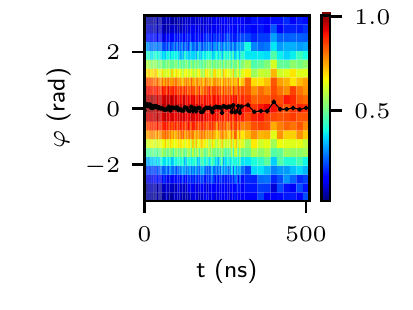}
			\end{minipage}
			\begin{minipage}[t]{0.25\linewidth}
				\textbf{$Q_4$}
		       	\centering
		       	\includegraphics[width=\linewidth]{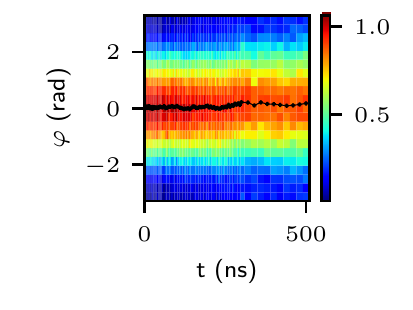}
          \end{minipage}
			\begin{minipage}[t]{0.25\linewidth}
				\textbf{$Q_5$}
				\centering
				\includegraphics[width=\linewidth]{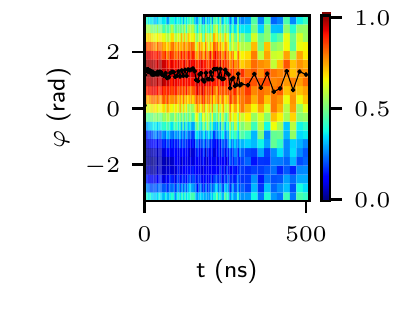}
			\end{minipage}
		}				
		\caption{\textbf{Experimental results of dynamical phase calibration.}  ~Black dots correspond to $\varphi = \phi _d $.  }\label{DyPhaseDeterminationResult}
	\end{figure}

\section{Supplementary Note 2. Wannier-Stark Localization Length}
Here, by Jordan-Wigger transformation, the Hamiltonian (2) can be mapped to a free-fermion lattice model
\begin{align} \label{Hfer}
 \hat H_{\textrm{eff}}=\sum_{\langle ij\rangle}g_{ij}(\hat{c}^\dagger_{i} \hat{c}_{j}+\hat{c}^\dagger_{j} \hat{c}_{i})+\sum_{j=1}^5h_{j}\hat{c}^\dagger_{j}\hat{c}_{j} ,
 \end{align}
where $\hat{c}_{j}^\dagger$ ($\hat{c}_{j}$) is the fermion creation (annihilation) operator.
Here, without loss of  generality, we consider $g_{ij}$ is site-independent, i.e., $g_{ij}=g$.
Under a linear potential, i.e., $h_j=Fj$, the system can exhibit Wannier-Stark localization with localization length $\xi_{WS}=2g/F$~\cite{Dahan1996}.
This localization length can be understood classically as following~\cite{Dahan1996}:
For a single particle, it can have the maximum kinetic energy $E_{max}^{K}=\max \big ( 2g\cos(k)\big)=2g$.
In addition, the particle need to  consume the kinetic energy $h$  when hopping to the next site due to the linear potential.
Thus, the maximum distance that a particle can propagate is about $2g/h$, i.e.,  Wannier-Stark localization length.

For a finite-size system, similarly, due to the linear potential, the particle cannot propagate completely from one boundary to another.
In this picture,  we can find that the maximum probability of a particle propagating from one boundary to another (i.e., $P_5^{max} $ in the main text.) can indeed reflect the localization length.
We assume that the localized wavefunction has the form
\begin{align} \label{psin}
\ket{\Psi_n}\sim A\sum_j^L e^{-|j-n|/\xi_{WS}}\hat c_j ^\dagger\ket{vac},
 \end{align}
where $A$ is a normalized factor, $n$ represents that the wave function is localized at site $n$, $\ket{vac}$ is the vacuum state of $\hat c_j$.
Therefore, intuitively, we expect  that the maximum probability of a particle propagating from one boundary to another is proportional to $e^{-\alpha/\xi_{WS}}$, where $\alpha$ is the corresponding factor.
In Supplementary Figure~\ref{fig_wsll}, we present the numerical result of the relation between $g/F$ and $P_5^{max}$, and
we can find that $P_5^{max} \propto g/F \propto1/\xi_{WS}$ with $F/g>0.6$.

\begin{figure}[t]     	
	\includegraphics[width=0.4\textwidth]{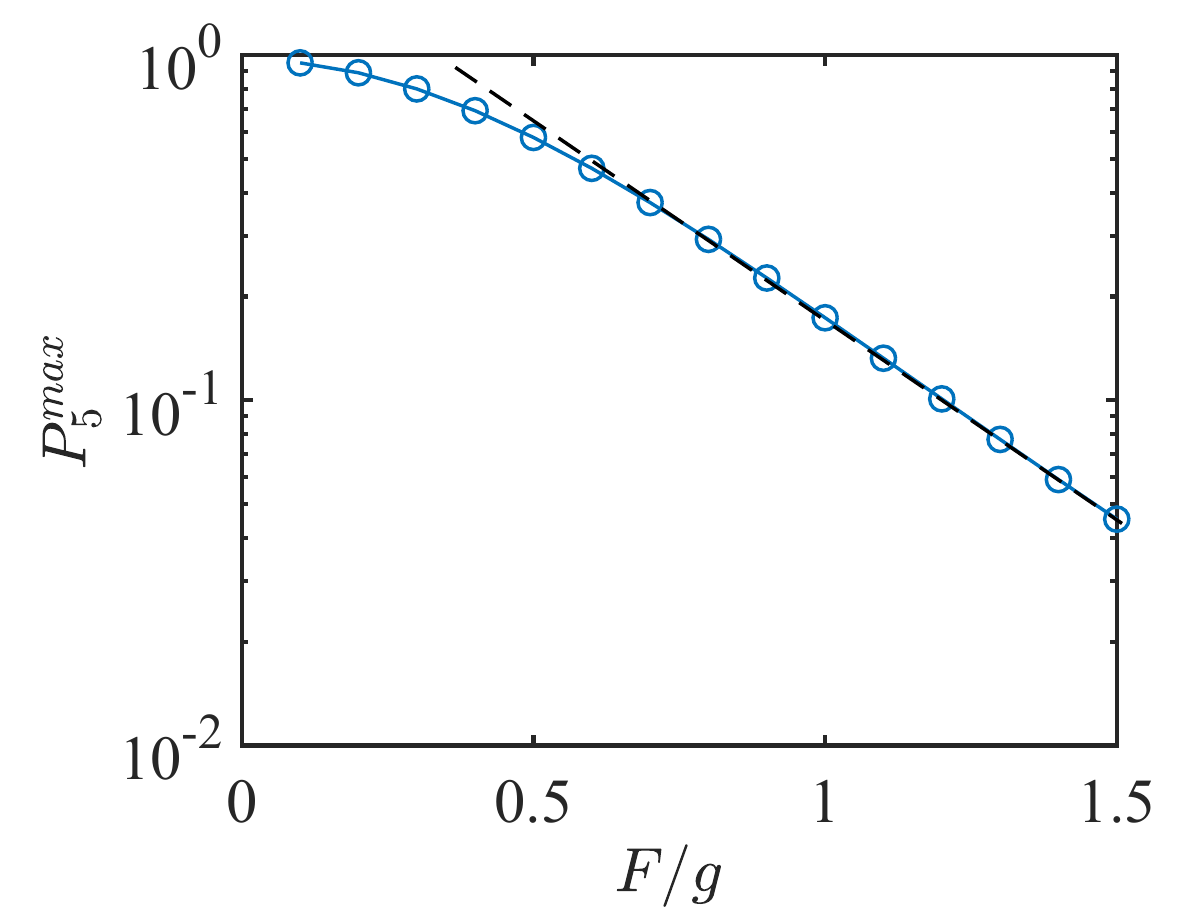}
	\caption{The relation between $g/F$ and $P^{max}_5$. Here we consider $g_{ij}=g=1$, and the initial state $\ket{10000}$.}
\label{fig_wsll}
\end{figure}

\section{Supplementary Note 3. Spin Current}
In this section, we consider the spin current under the linear potential.
The spin current  of Hamiltonian (2) can be defined as
 \begin{align}
  \hat J_n :=i(\hat{\sigma}^+_{n} \hat{\sigma}^-_{n+1}-\hat{\sigma}^-_{n} \hat{\sigma}^+_{n+1})=\frac{1}{2}(\hat{\sigma}^x_{n} \hat{\sigma}^y_{n+1}-\hat{\sigma}^y_{n} \hat{\sigma}^x_{n+1}),
 \end{align}
 where $n$ represents the $n$-th bond of the chain.
It can also be measured in our platform by the joint readout of two nearest-neighbor qubits.
Here, we present the numerical results [see Supplementary Figure~\ref{spin_current}].
When $F=0$,  the spin current has nearly no decay as the increase of the propagation distance, while it decays quickly for $F\neq0$.
Thus, this behavior of spin currents is an another signature of Wannier-Stark localization.

\begin{figure}[t]     	
	\includegraphics[width=0.5\textwidth]{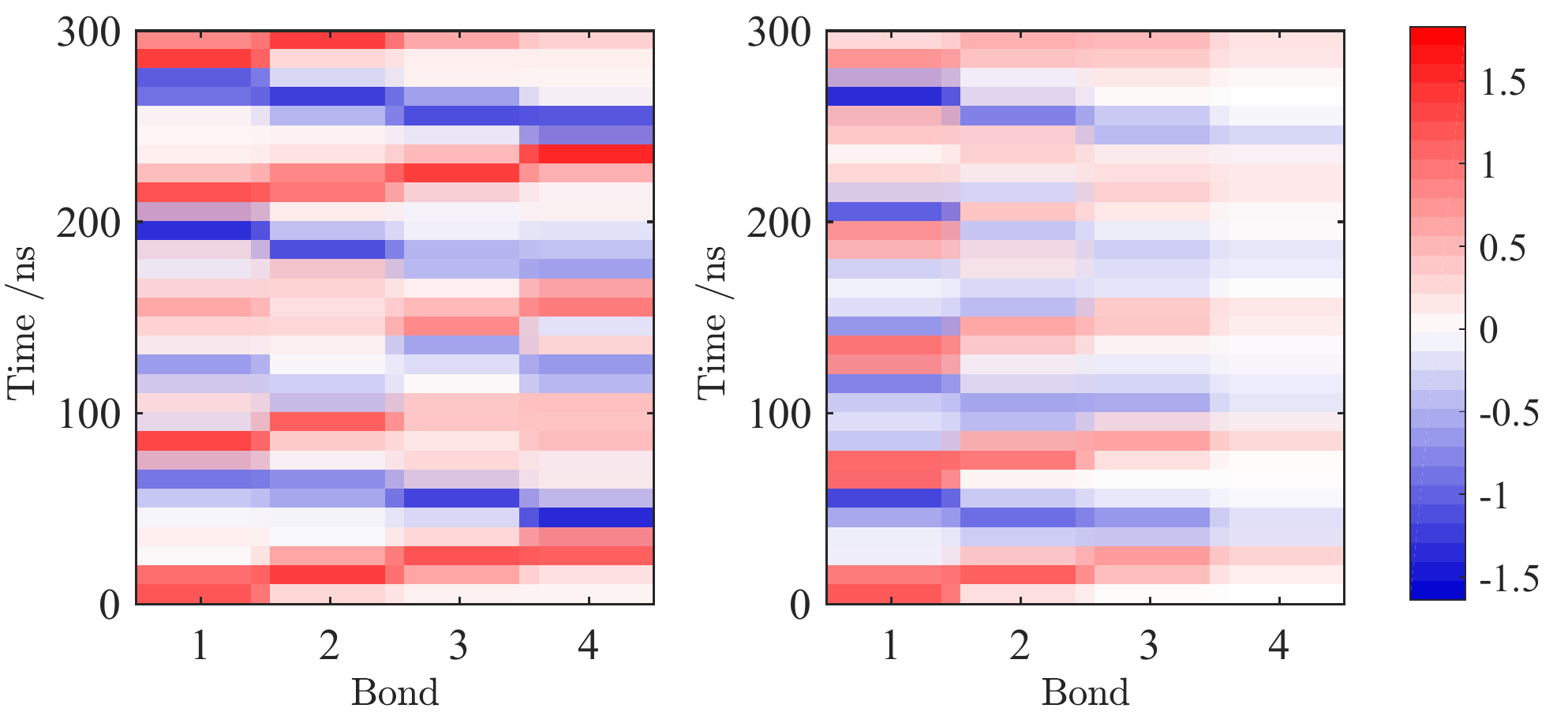}
	\caption{Numerical results of spin currents for $F/2\pi=0$ MHz (left) and $F/2\pi=10$ MHz (right). Here, the parameters of the Hamiltonian is from the device. The initial state is $\ket{10000}$. }
\label{spin_current}
\end{figure}

\end{document}